\documentclass[a4paper, 10pt]{article}
\usepackage{jheplike}

\usepackage{graphicx}              
\usepackage{amsmath}               
\usepackage{amsfonts}              
\usepackage{amsthm}   		         
\usepackage{color}	
\usepackage{psfrag}
\usepackage[all, knot]{xy}
\usepackage{calc}

\newcommand\nn{\nonumber}
\newcommand\be{\begin{eqnarray}}
\newcommand\ee{\end{eqnarray}}

\newcommand{\cA}{{\mathcal A}}
\newcommand{\cB}{{\mathcal B}}

\newcommand{\cD}{{\mathcal D}}

\newcommand{\cF}{{\mathcal F}}

\newcommand{\cH}{{\mathcal H}}

\newcommand{\cK}{{\mathcal K}}
\newcommand{\cL}{{\mathcal L}}

\newcommand{\cP}{{\mathcal P}}

\newcommand{\cT}{{\mathcal T}}

\newcommand{\cV}{{\mathcal V}}

\newcommand{\fc}{\mathfrak{c}}

\newcommand{\fh}{\mathfrak{h}}

\newcommand\U{{\mathrm U}}
\newcommand\C{{\mathbb C}}
\newcommand\N{{\mathbb N}}
\newcommand\R{{\mathbb R}}

\newcommand\K{{\mathcal{K}}}

\newcommand{\SU}{{\rm SU}}

\newcommand{\su}{\mathfrak{su}}

\newcommand\Tr{{\rm Tr}}

\newcommand\Hom{{\rm Hom}}

\newcommand{\braaket}[2]{\langle #1  \mid   #2 \rangle }
\newcommand{\ketbra}[2]{  \mid   #1 \rangle \langle #1 \mid}
\newcommand{\braaaket}[3]{\langle #1  \mid   #2   \mid   #3 \rangle }
\newcommand{\bra}[1]{\langle #1  |}
\newcommand{\ket}[1]{| #1 \rangle}
\newcommand{\ketr}[1]{| #1 ]}
\newcommand{\brar}[1]{[ #1 |}

\newcommand\bz{\bar{z}}

\newcommand{\braaketrr}[2]{[ #1  \mid   #2 ] }
\newcommand{\braaketrt}[2]{[ #1  \mid   #2 \rangle }
\newcommand{\braakettr}[2]{\langle #1  \mid   #2 ] }
\newcommand\one{\mathbb{I}}
\newcommand\norm[1]{\| #1 \|}

\newcommand{\V}{\mathcal{V}}
\newcommand{\hf}{\hat{f}}

\newcommand{\tz}{\tilde{z}}

\newcommand{\vectwo}[2]
{
   \begin{pmatrix} #1 \\ #2 \end{pmatrix}
}

\begin{document}

\title{Spinor Representation for Loop Quantum Gravity}
\author{Etera Livine,}
\author{Johannes Tambornino}
\affiliation{Laboratoire de Physique, ENS Lyon,\\ CNRS-UMR 5672,\\ 46 All\'ee d'Italie, Lyon 69007, France}
\emailAdd{etera.livine@ens-lyon.fr, johannes.tambornino@ens-lyon.fr}

\abstract{We perform a quantization of the loop gravity phase space purely in terms of spinorial variables, which have recently been shown to provide a direct link between spin network states and simplicial geometries. The natural Hilbert space to represent these spinors is the Bargmann space of holomorphic square-integrable functions over complex numbers. We show the unitary equivalence between the resulting generalized Bargmann space and the standard loop quantum gravity Hilbert space by explicitly constructing the unitary map. The latter maps $\SU(2)$-holonomies, when written as a function of spinors, to their holomorphic part. We analyze the properties of this map in detail. We show that the subspace of gauge invariant states can be characterized particularly easy in this representation of loop gravity. Furthermore, this map provides a tool to efficiently calculate physical quantities since integrals over the group are exchanged for straightforward integrals over the complex plane.
 }

\maketitle


\section{Introduction}
Loop quantum gravity (for reviews see \cite{thiemannbook, rovellibook, ashtekar_lewandowski_review}) and spin foam models (for reviews see \cite{perez_spinfoams, baez_spinfoams, daniele_thesis}) are attempts to construct a quantum theory of gravity, i.e. a framework to describe the behavior of gravitational dynamics at the Planck scale. The former is based on a reformulation of classical general relativity in terms of Ashtekar's variables and a subsequent application of rigorous canonical quantization methods carefully keeping track of general relativity's background independence at each step. The latter is a path integral approach to quantum gravity based on a reformulation of general relativity as an almost-topological field theory but also is related to discrete approaches to quantum gravity such as Regge-calculus due to the interpretation of spinfoams as (quantized) polyhedral decompositions of the spacetime manifold. Remarkably, despite having such different starting points, the two frameworks have converged in the last years, at least on the kinematical level\footnote{For full general relativity in four spacetime dimensions a rigorous link between  spinfoam models and canonical loop quantum gravity on the \emph{dynamical} level is still missing. However, there exist strong hints that this equivalence also holds at the dynamical level, see for example \cite{noui_perez_04} for an affirmative answer to that question in three dimensional gravity and \cite{valentin_hamiltonian, valentin_laurent} for interesting recent developments in the context of three- and four-dimensional $BF$-theory.}, i.e. without taking the Hamiltonian constraint into account: the theory's state space (or boundary state space from the spin foam point of view) is made of the so called spin network functions, which are gauge-invariant functions on (multiple copies of) the group $\SU(2)$ that live on graphs. These graphs directly define three-dimensional space, the geometry of space is entirely encoded in the spin network functions living on that graph.\\
The loop quantum gravity approach on the one hand is particularly successful in providing a strong link of this theory with classical general relativity, also because the question of the continuum limit can be answered in an elegant and convincing way due to use of so called projective techniques (see for example \cite{ashtekar_lewandowski_94_1}). However, the definition of dynamics in the canonical framework is notoriously complicated\footnote{There exist a mathematically well defined quantization of the Hamiltonian constraint \cite{thiemann_hamiltonian} due to Thiemann since roughly 15 years, but due to its complexity to date it is still unclear whether this definition of dynamics is similar to that of general relativity; for an alternative proposal see also \cite{alesci_hamiltonian} }, one main reason being that from this point of view the geometrical picture behind the spin network functions remains relatively indirect. The spinfoam approach on the other hand, whose continuum limit towards classical gravity is less understood, is very strong in providing exactly this geometric interpretation because spin network functions can be interpreted as a quantization of classical discrete chunks of space. This allowed for a new route for the definition of dynamics \cite{epr1, epr2, eprl, fk}. See \cite{maite_thesis} for a pedagogical review of the latest results.\\
A particularly appealing way to understand the geometry behind spin network states, which was developed from two slightly different perspectives in \cite{laurentsimone, laurentsimone2} and \cite{laurent_etera_un1, laurent_etera_un2} and subsequently build upon in \cite{etera_spanish_un1, etera_spanish_un2, etera_spanish_laurent, dupuis_livine_intertwiner, dupuis_livine_intertwiner2} is the following: the Hilbert space associated to one single edge of a graph in loop quantum gravity is given by $\cH_e := L^2(\SU(2), dg)$, the space of square integrable functions over the group $\SU(2)$. This can be seen as the quantization of a classical phase space $T^*\SU(2) \simeq \SU(2) \times \su(2)$, which is simply the cotangent bundle over $\SU(2)$. Typically, the tuple $(g,X)$ consisting of a group element $g$ and a Lie algebra element $X$ is used as a basis on that space. However, one can also choose different coordinate systems on $T^*\SU(2)$. In the works cited above it became clear that one particular convenient choice of such coordinate system is given by two $\C^2$-spinors $(\ket{z}, \ket{\tz})$ and it was shown that the classical physics of the phase space $\cT^*\SU(2)$ can entirely be formulated in terms of these spinors.\\
The big advantage of the spinor formalism is its very natural link to (discrete) geometry: Each spinor can be seen to define a vector in $\R^3$ (up to $\U(1)$) and these vectors can in turn be interpreted as providing the area vectors of the faces of elementary polyhedra. Thus, the classical phase space $[T^*\SU(2)]^E$ associated to the Hilbert space of a given graph $\gamma$ with $E$ edges in loop gravity can be understood as a  space of polyhedra glued together in an appropriate way to provide a piecewise flat manifold (see \cite{simone_eugenio} for a detailed analysis of the classical polyhedral phase space). The curvature of this manifold is then described by the way the individual polyhedra are glued together.\\
The spinorial formalism opens a new route towards understanding the quantum geometry of loop  gravity which we develop in detail in this article: Considering the spinorial variables as fundamental, and group elements and Lie algebra elements as composite, one is led to a different quantization of $T^*\SU(2)$ where the Hilbert space $\cH^{\rm spin}_e$ is given by an appropriate gauge-reduction (implementing the condition that the spinors on both sides of the edge have the same norm) of the Bargmann space of holomorphic square-integrable functions in both spinors. Elementary operators on this space are the ladder operators associated to the spinors, from which flux- and holonomy-operators can be derived as composite operators.\\
This Hilbert space and the space $\cH_e$ used in standard loop quantum gravity arise as quantizations of the same space using different polarizations. Since different polarizations generally lead to different quantizations there is a priori no reason to believe that the quantization in terms of spinors captures the same physics as the quantization in terms of group variables, i.e. that the two Hilbert spaces are unitarily equivalent. However, we construct a map $\cT: \cH_e \rightarrow \cH_e^{\rm spin}$ that maps group representation matrices of $\SU(2)$ onto holomorphic functions in $(\ket{z}, \ket{\tz})$ that can easily be seen to be unitary. From an abstract point of view this map can be understood as the restriction to the holomorphic part of the group element $g$ written in terms of spinors\footnote{The spirit of the construction of our unitary transform bears some similarity with the one of Hall \cite{hall} who constructed a unitary map between $L^2(\SU(2), dg)$ and $L^2_{\rm hol}(\SU(2)^\C, d\mu)$, the space of holomorphic square-integrable functions over the complexification of $\SU(2)$ with a heat kernel measure. This transform was generalized to the continuum context of loop quantum gravity in \cite{almmt_coherent}. However, since functions that are holomorphic in the spinors are not necessarily holomorphic functions in the sense of Hall (due to different choices of complex polarization),these seem not to be the same.}. This map can trivially be lifted to the Hilbert space $\cH_\gamma$ associated to a fixed graph in loop quantum gravity, and furthermore is compatible with the inductive limit construction performed in loop quantum gravity to generate the continuum theory Hilbert space $\cH$ by appropriately glueing together the Hilbert spaces $\cH_\gamma$ of \emph{all possible} graphs. Thus, all loop quantum gravity can be mapped onto \emph{holomorphic functions} of certain spinor variables in a unitary way.\\
Apart from the obvious advantage of a clear geometric interpretation of these states, this formalism has one further big advantage to standard loop gravity: the fundamental variables are now spinors and not group variables anymore, which will simplify calculations a lot. As integrals over $\SU(2)$ are exchanged for straightforward integrals over the complex plane, closely related to the moments of a Gaussian measure, we expect that computations of physical quantities, such as correlation function in the spinfoam formalism, will substantially simplify using the spinor-representation of loop quantum gravity. As an example, we show that the Haar measure on $\SU(2)$, written in terms of spinors, is simply given by the product of two standard Gaussian measures, which illustrates the simplifications for practical calculations that are expected to occur.\\
Further, using the holomorphic variables simplifies the construction of $\SU(2)$-invariant states, which is well understood in loop quantum gravity but rather ugly from the practical point of view, as one needs to use the machinery of $\SU(2)$-recoupling theory. Extending the results of \cite{laurent_etera_un2, etera_spanish_laurent} to our unitary map leads to a simple characterization of $\SU(2)$-invariant spin networks in terms of holomorphic functions of the spinors.
\vspace{1cm}\\
The article is organized as follows: in section \ref{spinor_variables} we briefly review the reformulation of the loop gravity phase space\footnote{By \emph{loop gravity phase space} we mean the classical phase space $[T^*\SU(2)]^E$ associated to a given graph $\gamma$ with $E$ edges. The Hilbert space $\cH_\gamma$ can then be seen as a quantization of that space.} written in terms of spinors. We introduce the spinorial variables from an abstract point of view and then explain how they constitute the loop gravity phase space. As a side-effect, we prove that the Haar measure on $\SU(2)$ is simply given by the product of two Gaussian measures when writing the group element in terms of spinors.\\
In section \ref{quantum_spinors} we restrict to a graph consisting of one single edge only to keep things simple: we take the point of view that the spinors are to be considered as fundamental variables and base our quantization of $T^*\SU(2)$ thereon. We introduce the Bargmann space of holomorphic square-integrable functions and construct the spinor Hilbert space $\cH^{\rm spin}_e$ by tensoring two Bargmann spaces together and solving the $\U(1)$-constraint. Then we construct a map $\cT: \cH_e \rightarrow \cH^{\rm spin}_e$ which can be understood as the holomorphic restriction of group representation matrices written in terms of spinors. This map can easily be seen to be unitary. We analyze the properties of this map, construct operators on $\cH^{\rm spin}_e$ and illustrate its action on some simple examples.\\
Then we leave the realm of one single edge and generalize the unitary map to an arbitrary graph in section \ref{sec:quantum_arbitrary}. This generalization is straightforward and establishes unitary equivalence of the loop quantum gravity Hilbert space $\cH_\gamma$ and the spinor Hilbert space $\cH^{\rm spin}_\gamma$. A bit more care is necessary to understand the continuum generalization of this map: however, as unitary equivalence holds on each graph separately it is not difficult to show that the conditions of cylindrical consistency are fulfilled on the spinor side as well when appropriately defining an equivalence relation between spinor states living on different graphs. Thus, the continuum limit exists on the spinor side and takes the same form as in standard loop quantum gravity. However, an intrinsic definition of that space is not available at the moment. Furthermore we carefully analyze the spinor Hilbert space from the point of view of $\SU(2)$-invariance: first we consider some simple examples (the one-loop graph and the two-vertex graph) to get some intuition before we construct gauge invariant spinor states for an arbitrary graph. These turn out to be related to the $\U(N)$-coherent states of \cite{laurent_etera_un2} but require some extra condition.\\
We supplement this article with some extra material to make it self-contained: In appendix \ref{segal_bargmann_appendix} we review the standard Segal-Bargmann transform as our unitary map can be seen as a generalization of the latter. In appendix \ref{coherent_state_appendix} we collect some definitions and formulae concerning the coherent state basis of $\SU(2)$ which we used in this article. In appendix \ref{lqg} we give a very short summary on the projective techniques and the conditions of cylindrical consistency used in loop quantum gravity as the construction of the continuum spinor Hilbert space in section \ref{sec:quantum_arbitrary} uses the same logic. Finally, in appendix \ref{factor_dj_appendix} we discuss a family of alternative measures which could be used instead of the Gaussian one on the spinor space and remark on some combinatorial factors that appear in our unitary map.

\section{LQG phase space in terms of spinors} \label{spinor_variables}
\subsection{The LQG phase space}
The (gauge-variant) kinematical Hilbert space used in loop quantum gravity $\cH := \overline{\cup_\gamma \cH_\gamma / \sim}$ is the (completion in an appropriate norm of the) union of certain Hilbert spaces $\cH_\gamma$ associated to graphs $\gamma$. Each of these Hilbert spaces is of the form $\cH_\gamma := L^2(\SU(2)^E, d^Eg)$ where $E$ is the number of edges of $\gamma$, each edge carries one copy of $\SU(2)$ and $d^Eg$ is the product Haar-measure on $\SU(2)^E$. Regardless its original derivation\footnote{See, for example, \cite{thiemannbook} for a complete account of the original derivation of this space as a quantization of classical general relativity written in terms of Ashtekar's variables. The detailed sense in which the ``union over all graphs'' has to be understood is strongly linked with this interpretation. See also appendix \ref{lqg} where we briefly collect some details about this space which are useful for the constructions carried out in this paper.}, abstractly each Hilbert space $\cH_\gamma$ can be interpreted as a quantization of the classical space $[T^*\SU(2)]^E \simeq [\SU(2) \times \su(2)]^E$, the cotangent bundle over $\SU(2)$. Typically, this cotangent bundle is parameterized in terms of a group element $g$ and a Lie-algebra variable $X = \vec{X}\cdot \vec{\sigma}$ where $\sigma^i, i=1,2,3$ are the antisymmetric Pauli matrices chosen Hermitian, traceless and normalized as $\Tr(\sigma^i \sigma^j) = \delta^{ij}$ and the symplectic structure is given by
\be \label{PB_SU(2)}
\{ g_{IJ}, g_{KL}  \} & = & 0 \nn \; ,\\
\{ X^i, X^j \} & = & \epsilon^{ij}_k X^k \nn \; , \\
\{ X^i , g_{IJ} \} & = & -\sigma^i g_{IJ} \; .
\ee
However, $(g,X)$ is not the only coordinate chart on $T^*\SU(2)$, for different purposes it might be useful to use a different parameterization of that space. From the point of view of interpreting spin network states in terms of simplicial geometries it has turned out to be especially useful to introduce a set of $\C^2$-spinors $(\ket{z}, \ket{\tz})$ as a coordinate chart on $T^*\SU(2)$. This framework was constructed from two slightly different perspectives in \cite{laurentsimone, laurentsimone2} and \cite{laurent_etera_un1, laurent_etera_un2} and subsequently build upon in \cite{etera_spanish_un1, etera_spanish_un2, etera_spanish_laurent} and converged to the following picture:\\
Instead of associating a group element $g$ and a Lie-algebra element $X$ to each edge one can equally well associate a pair of \emph{spinors} $(\ket{z}, \ket{\tz})$ to the initial and the final vertex of that edge respectively. The standard symplectic structure on $\C^2 \times \C^2$ turns this space into a phase space. The group elements $g(z,\tz)$ and Lie-algebra element $X(z,\tz)$ then emerge as composite functions constructed out of these spinors, and one obtains back the loop gravity phase space for one edge with symplectic structure (\ref{PB_SU(2)}) after a gauge reduction by $\U(1)$ which demands that the two spinors on opposite sides of the same edge have equal length. This symplectomorphism was proven and analyzed in detail in \cite{laurentsimone}.\\
However, keeping in mind that $\cH_{e}$ is only the \emph{gauge-variant} edge Hilbert space of loop quantum gravity and the full kinematical data is only obtained after appropriately glueing together these edge Hilbert spaces such that $\SU(2)$-invariance at each vertex is achieved, there is a second route one can follow: although well understood in principle, the construction of these \emph{intertwiner spaces} that capture the $\SU(2)$-invariant information of the spin network states is a rather cumbersome task, because one has to dive into $\SU(2)$-recoupling theory. In \cite{laurent_etera_un1, laurent_etera_un2, etera_spanish_laurent} it was observed that from the point of view of the spinor framework it is very easy to construct classical $\SU(2)$-invariant quantities at each vertex separately -- simply because the spinors are not located on the edges, as the group variables, but truly live on the vertices. These $\SU(2)$-invariant quantities are simply given by gauge-invariant combinations of spinors living at the same vertex, such as $E_{ij} := \braaket{z_i}{z_j}$. Glueing these matrices living at different vertices together in a $\U(1)$-invariant fashion one obtains the classical phase space corresponding to $\SU(2)$-invariant spin network functions. In the following we will briefly explain this spinor formalism and its interpretation as an extension (with an additional $\U(1)$-symmetry) of the loop gravity phase space.

\begin{figure}
\be
\xymatrix{
f(z,\tz)  \ar[d]_{\SU(2)} \ar[r]^{\U(1)} & **[r]  f(g)   \ar[d]^{\SU(2)}
  \\
 f(F,E)  \ar[r]^{\U(1)}   & **[r] f(F,E)|_{\U(1)} \simeq f(g)|_{\SU(2)}
} \nn
\ee
\caption{When considering the spinors $\ket{z}$ as fundamental variables one can follow different paths: on the one hand one can first divide out the $\U(1)$-gauge invariance on the edges, which leads back to the standard phase space of loop gravity with group variables $g$ and Lie-algebra variables $X$. Then one still has to divide out by $\SU(2)$. On the other hand one can also first divide out the $\SU(2)$-gauge invariance at the vertices, which leads to the $\U(N)$-framework and a characterization of the intertwiner-space in terms of $E$- and $F$-variables. These spaces then have to be glued together in an appropriate way to respect $\U(1)$-gauge invariance at the edges. }
\end{figure}

\subsection{Spinor variables}
Denote by $\ket{z} \in \C^2$ and $\bra{z} \in \C^2$ a spinor and its conjugate respectively, i.e.
\be
\ket{z} := \vectwo{z^0}{z^1}, \qquad \bra{z} := (\bz^0, \bz^1) \; .\nn
\ee
Being elements of $\C^2$ the spinors transform naturally under the defining representation of $\SU(2)$,
\be
h: \C^2 \rightarrow \C^2; \; \ket{z} \mapsto h\ket{z} \; \forall h \in \SU(2) \;.\nn
\ee
Endow $\C^2$  with the standard, positive inner product $\braaket{w}{z} := \bar{w}^0z^0 + \bar{w}^1z^1$ and a norm $\norm{z} := \sqrt{ \braaket{z}{z} }$.
Further, using the antisymmetric matrix
\be
\epsilon := \begin{pmatrix} 0 & -1 \\ 1 & 0\end{pmatrix}\nn
\ee
we define an anti-unitary map as
\be
\ket{z} \mapsto \ketr{z} := \epsilon \ket{\bar{z}} \; .\nn
\ee
The object $\ketbra{z}{z}$ is a hermitian 2x2 matrix and, thus, can be decomposed in a basis $(\one, \sigma^i)$, where $\sigma^i, i=1,2, 3$ are the hermitian traceless generators of $\su(2)$, normalized as $tr(\sigma^i \sigma^j) = \delta^{ij}$:
\be \label{zandX}
\ketbra{z}{z} = \frac{1}{2}(\braaket{z}{z}\one + \vec{X}\cdot \vec{\sigma}) \; .
\ee
Thus, a spinor defines a three dimensional vector $\vec{X}$ whose norm is $|\vec{X}| = \braaket{z}{z}$ and whose components are explicitly given by
\be \label{X_z}
X^1 = 2\Re(z^0 \bz^1), \; X^2 = 2\Im(z^0 \bz^1), \; X^3 = |z^0|^2 - |z^1|^2 \; .
\ee
However, the map is not one-to-one as the $\U(1)$-transformation $\ket{z} \stackrel{\U(1)}{\mapsto} e^{i\theta} \ket{z}$ leaves the vector $\vec{X}$ invariant. This has to be kept in mind when relating this formalism to loop gravity and extracting physical information.\\
Now assume that we have two pairs of such spinors, $\ket{z}$ and $\ket{\tz}$, sitting at the beginning and final vertex of an edge respectively. It is possible to construct an $\SU(2)$-group element $g(z, \tz)$ out of these two spinors as
\be \label{group_element}
g(z, \tz) := \frac{  \ket{z}\brar{\tz} - \ketr{z}\bra{\tz} }{\norm{z} \norm{\tz}} \; ,
\ee
$g(z, \tz)$ can easily be checked to be a proper element of $\SU(2)$ and its inverse is given by
\be
g^{-1}(z, \tz) := \frac{  -\ket{\tz}\brar{z} + \ketr{\tz}\bra{z} }{\norm{z} \norm{\tz}} \, .
\ee
Therefore the following property holds:
\be
g\frac{\ket{\tz}}{\norm{\tz}} = -\frac{\ketr{z}}{\norm{z}} \quad g^{-1}\frac{\ket{z}}{\norm{z}} =  \frac{\ketr{\tz}}{\norm{\tz}} \; . \nn
\ee
Thus it rotates the (normalized) spinor $\ket{z}$ into the (normalized) dual spinor $\ketr{\tz}$ and vice versa (up to a sign).\\
Furthermore, under local $\SU(2)$-transformations $g(z, \tz)$ transforms exactly as the holonomy of an $\SU(2)$-connection, i.e.
\be
(\ket{z}, \ket{\tz}) & \stackrel{\SU(2)}{\mapsto} & (h_1\ket{z}, h_2\ket{\tz}) \nn \\
& \mbox{implies}& \nn \\
g(z, \tz) & \stackrel{\SU(2)}{\mapsto} & h_1 g(z, \tz) h_2^{-1} \; \forall h_1, h_2 \in \SU(2) \; .\nn
\ee
When looking at the $\su(2)$-elements $X(z) := \vec{X}(z) \cdot \vec{\sigma}$ and $\tilde{X}(\tz) := \vec{\tilde{X}}(\tz) \cdot \vec{\sigma}$ one realizes that these fulfill the identity
\be \label{X_id}
\tilde{X} = -g^{-1}Xg \;
\ee

\subsection*{Spinorial phase space}
Now equip the space $\C^2 \times  \C^2$, spanned by the two spinors $\ket{z}, \ket{\tz}$ with a symplectic structure and, thus, turn it into a phase space:
\be
\{ \bz^i, z^j  \} = i \delta^{ij} \quad  \{\bar{\tz}^i, \tz^j   \} = i \delta^{ij} \; ,\nn
\ee
where $i,j = 0,1$. This is the standard complex structure on $\C^4$.\\
On $\C^2 \times \C^2$ consider a constraint
\be \label{u1_constraint}
\fh := \norm{z}^2 - \norm{\tz}^2
\ee
which just states that the two spinors $z, \tz$ should have the same length. It is easy to see that $\fh$ generates $\U(1)$-gauge transformations:
\be
\{ \fh, \ket{z} \} = i\ket{z} \quad \{ \fh, \ket{\tz} \} = -i\ket{\tz} \nn \; .
\ee
Thus, finite gauge transformations are given by
\be
\ket{z} \stackrel{\U(1)}{\mapsto}  e^{i\theta} \ket{z} \qquad \ket{\tz}  \stackrel{\U(1)}{\mapsto}   e^{-i\theta} \ket{\tz} \; .\nn
\ee
It is immediate to see that the group- and Lie-algebra elements are indeed $\U(1)$-invariant,
\be
g \stackrel{\U(1)}{\mapsto} g \qquad X \stackrel{\U(1)}{\mapsto} X \; .\nn
\ee
A non-trivial result, established in \cite{laurentsimone2}, is that starting from the spinor space $\C^2 \times \C^2$ one can obtain the cotangent bundle of $\SU(2)$ by $\U(1)$-gauge reduction\footnote{To be precise, gauge reduction has to be understood in the sense of the ``double quotient'' space, i.e. those elements of $\C^2 \times \C^2$ which (i) lie on the constraint surface of $\fh$, and (ii) which Poisson-commute with $\fh$. The symplectomorphism exists only for elements $(g,X) \in T^*\SU(2)$ with $|X| \neq 0$.}:
\be
\C^2 \times \C^2 \backslash \{ \braaket{z}{z}=0, \braaket{\tz}{\tz}=0 \} // \U(1)  \simeq T^*\SU(2) \backslash \{ |X|= 0 \} \, .\nn
\ee
In particular, starting from the natural symplectic structure on $\C^2 \times \C^2$ one can derive the deduced symplectic structure for $g$ and $X$ which (on the constraint hypersurface, i.e. when $\fh=0$ is satisfied) turns out to be
\be
\{g_{IJ}(z, \tz), g_{KL}(z, \tz)   \} &  = & 0 \; , \nn \\
\{X^i(z), X^j(z)\} & = & \epsilon^{ij}_k X^k(z) \; , \nn \\
\{X^i(z), g_{IJ}(z, \tz)   \} &  = & -\sigma^i  g_{IJ}(z, \tz) \; , \nn \\
\{\tilde{X}^i(\tz), g_{IJ}(z, \tz)\} & = &  g_{IJ}(z, \tz)\sigma^i \; .\nn
\ee
Thus, together with relation (\ref{X_id}) this is identical to the standard symplectic structure on $\SU(2)$ (\ref{PB_SU(2)}).\\
Applied to loop gravity this gives an interesting picture: consider a graph $\gamma$ consisting simply of one edge $e$ between vertex $v_1$ and vertex $v_2$. Then, instead of assigning a group element $g$ and a Lie-algebra element $X$ to the edge, one can equally well assign doublet of spinors $\ket{z}, \ket{\tz}$ to the vertices. Thus, the dynamical degrees of freedom are shifted to the vertices of the graph in this interpretation.

\subsection{The Haar measure on $\SU(2)$ in terms of spinors} \label{haar_measure}
Considering the spinors as fundamental quantities and the group elements as composite ones also allows a different perspective on the Haar measure $dg$ itself: writing $g(z, \tz)$ in terms of spinors can be seen as choosing a particular coordinate system on $\SU(2)$ with a lot of redundant degrees of freedom (8 real degrees of freedom in $(\ket{z}, \ket{\tz})$ compared to 3 real degrees of freedom in $g$). $g$ does not depend on the length of the two spinors used (which reflects the $\U(1)$-invariance discussed earlier) and is further invariant under a ``twisted'' rotation of both spinors by an arbitrary element $h \in \SU(2)$  as
\be
\vectwo{\ket{z}}{\ket{\tz}} \stackrel{h}{\mapsto} \vectwo{h\ket{z}}{g^{-1}hg\ket{\tz}}\, . \nn
\ee
The straightforward $\SU(2)$ action, $z\rightarrow h z,\, \tz\rightarrow h \tz$, would not leave 
the group element $g$ invariant and sends it to $hgh^{-1}$. The twisted rotation defined above allows to correct this and shifts the two spinors while leaving $g$ invariant.

What does the Haar measure $dg$ look like in these coordinates? Surprisingly, when using the spinors to parameterize $\SU(2)$ the Haar measure is just given by a normalized Gaussian on $\C^4$ in the sense that
\be
\int\limits_{\SU(2)} dg f(g) = \int\limits_{\C^2 \times \C^2} d\mu(z) d\mu(\tz) f(g(z,\tz)) \, , \nn
\ee
where $d\mu(z) := \frac{1}{\pi^2}dz^0 dz^1 e^{-\braaket{z}{z} } $ is the normalized Gaussian measure on $\C^2$ and the parameterization $g(z,\tz)$ is understood as in (\ref{group_element}). This can be proven by calculating the scalar product between two representation matrices of $\SU(2)$, written in terms of spinors. To simplify the computations we use the representation `matrices' $D^j_{\omega \tilde{\omega}}(g)$ in the coherent state basis which are labeled by two spinors $\ket{\omega}, \ket{\tilde{\omega}} \in \C^2$ and are given explicitely by  (see appendix \ref{coherent_state_appendix} for their relation to the more commonly used Wigner matrix elements in the magnetic number basis)
\be
D^j_{\omega \tilde{\omega}}(g) = \braaaket{\omega}{g}{\tilde{\omega}}^{2j} \, . \nn
\ee
The final result is
\be \label{gleichung_1}
\int d\mu(z) d\mu(\tz) \overline{D^j_{\omega \tilde{\omega}}(g(z, \tz))} D^k_{\alpha \tilde{\alpha}}(g(z,\tz)) & = & \frac{\delta^{jk}}{d_j}\braaket{\alpha}{\omega}^{2j} \braaket{\tilde{\omega}}{\tilde{\alpha}}^{2j} \, ,
\ee
as expected for the scalar product between 2 representation matrices in the coherent state basis. To prove the last equality, we will use an alternative, more symmetric, splitting of the group element $g$ into
\be
g(z, \tz) := \frac{\ket{z}\bra{\tz} + \ketr{z}\brar{\tz}}{\sqrt{\braaket{z}{z} \braaket{\tz}{\tz}}} \, .\nn
\ee
This definition is not invariant under the action of $\U(1)$ as defined before\footnote{Instead, this definition is invariant under simultaneous multiplication of both spinors with \emph{the same} phase, not the opposite one as (\ref{group_element}). The corresponding constraint generating this $\U(1)$-transformation would be (\ref{u1_constraint}) with positive sign.} but will simplify the combinatorics in the following proof a bit: we write the left hand side of (\ref{gleichung_1})  as
\be \label{formula_1}
\int d\mu(z) d\mu(\tz) \frac{\left[ \braaket{\tilde{\omega}}{\tilde{z}}\braaket{z}{\omega} + \braakettr{\tilde{\omega}}{\tz} \braaketrt{z}{\omega}  \right]^{2j}  \left[  \braaket{\alpha}{z}\braaket{\tilde{z}}{\tilde{\alpha}}  + \braakettr{\alpha}{z} \braaketrt{\tz}{\tilde{\alpha}}   \right]^{2k}}{\braaket{z}{z}^{j+k}\braaket{\tz}{\tz}^{j+k}} \, .
\ee
Apart from the norm factors in the denominator this integral can be computed using Wick's theorem, as this amounts to simply computing the moments of two Gaussian measures separately. Fortunately, introducing a pseudo-spherical coordinate system on $\C^2$, one can show that leaving away these norm factors in the above integral just amounts to multiplication by some combinatorial factor. Write
\be
\ket{z} = \vectwo{r \cos\theta e^{i \phi}}{r \sin \theta e^{i\psi}}\nn
\ee
where $r \in [ 0, \infty\}, \theta \in [ 0 , \pi/2 \}, \phi \in [ 0, 2 \pi \}, \psi \in [ 0, 2\pi \}$ and the Gaussian measure takes the form
\be
\int\limits_{\C^2}d\mu(z) = \frac{1}{\pi^2}\int\limits_0^\infty dr r^3 e^{-r^2} \int\limits_0^{\pi/2}d\theta \cos \theta \sin \theta \int\limits_0^{2\pi} d\phi  \int\limits_0^{2\pi} d\psi \, .\nn
\ee
Due to $\braaket{z}{z} = r^2$ it can easily be seen that the group element $g$ is independent of the radial coordinate, and thus the integration over $r$ in (\ref{formula_1}) decouples from the rest:
\be
\int\limits_0^\infty dr r^3 e^{-r^2} = \frac{1}{2} \, .\nn
\ee
Leaving away the norm factors in (\ref{formula_1}) then simply amounts to changing the last integral to
\be
\int\limits_{0}^\infty dr r^{3+2(j+k)} e^{-r^2} = C(j+k) \, ,\nn
\ee
where the combinatorial factor is given by
\be
C(j+k) = \begin{cases} \frac{1}{2}(j+k+1)! \quad \qquad \qquad \qquad \text{ if } $2(j+k)$ \text{even } \, , \\ \sqrt{\pi}(\frac{1}{2})^{2(j+k) + 4} \frac{(2(j+k)+3)!}{(j+k+\frac{3}{2})!}  \qquad \text{if } $2(j+k)$ \text{odd } \, .    \end{cases}\nn
\ee
Thus we can reduce (\ref{formula_1}) to a Gaussian integral over polynomials in $\ket{z}$ and $\ket{\tz}$ as
\be \label{formula_2}
& & \int d\mu(z) d\mu(\tz) \frac{\left[ \braaket{\tilde{\omega}}{\tilde{z}}\braaket{z}{\omega} + \braakettr{\tilde{\omega}}{\tz} \braaketrt{z}{\omega}  \right]^{2j}  \left[  \braaket{\alpha}{z}\braaket{\tilde{z}}{\tilde{\alpha}}  + \braakettr{\alpha}{z} \braaketrt{\tz}{\tilde{\alpha}}   \right]^{2k}}{\braaket{z}{z}^{j+k}\braaket{\tz}{\tz}^{j+k}} \nn \\
& = & \left[ \frac{C(0)}{C(j+k)} \right]^2 \int d\mu(z) d\mu(\tz) \left[ \braaket{\tilde{\omega}}{\tilde{z}}\braaket{z}{\omega} + \braakettr{\tilde{\omega}}{\tz} \braaketrt{z}{\omega}  \right]^{2j}  \left[  \braaket{\alpha}{z}\braaket{\tilde{z}}{\tilde{\alpha}}  + \braakettr{\alpha}{z} \braaketrt{\tz}{\tilde{\alpha}}   \right]^{2k} \nn \\
& = & \left[ \frac{C(0)}{C(j+k)} \right]^2  \sum\limits_{L= 0}^{2j} \sum\limits_{M=0}^{2k}\frac{(2j)!(2k)!}{(2j-L)!L!(2k-M)!M!} \nn \\
&& \qquad \qquad \int d\mu(z) \braaket{z}{\omega}^{2j-L}\braaket{\alpha}{z}^{2k-M}\braakettr{z}{\alpha}^M\braaketrt{\omega}{z}^L \times \nn \\
&& \qquad \qquad \times \int d\mu(\tz) \braaket{\tz}{\tilde{\alpha}}^{2k-M}\braaket{\tilde{\omega}}{\tz}^{2j-L}\braakettr{\tz}{\tilde{\omega}}^L\braaketrt{\tilde{\alpha}}{\tz}^M
\ee
These integrals can be evaluated separately using Wick's theorem, which states that the moments of a Gaussian integral are simply given by the sum over all possible pairings. Taking into account that $\braaketrt{\omega}{\omega}= 0$ one obtains
\be
\int d\mu(z) \braaket{z}{\omega}^{2j-L}\braaket{\alpha}{z}^{2k-M}\braakettr{z}{\alpha}^M\braaketrt{\omega}{z}^L & = & \delta^{jk}\delta_{LM} (2j-L)!L! \braaket{\alpha}{\omega}^{2j-L}\braaketrr{\omega}{\alpha}^{L} \nn \\
& = & \delta^{jk} \delta_{LM} (2j-L)!L!\braaket{\alpha}{\omega}^{2j} \, .\nn
\ee
Putting all this together and taking care of the combinatorial factor one obtains
\be
\int d\mu(z)d \mu(\tz) \overline{D^j_{\omega \tilde{\omega}}(g(z,\tz))} D^j_{\alpha \tilde{\alpha}}(g(z,\tz)) & = & \frac{\delta^{jk}}{[(2j+1)!]^2}[(2j)!]^2\braaket{\alpha}{\omega}^{2j} \braaket{\tilde{\omega}}{\tilde{\alpha}}^{2j} \sum\limits_{L=0}^{2j} 1 \nn \\
& = & \frac{\delta^{jk}}{d_j} \braaket{\alpha}{\omega}^{2j} \braaket{\tilde{\omega}}{\tilde{\alpha}}^{2j} \, ,\nn
\ee
which completes the proof.\\
It is important to note that we get \emph{exactly} the right combinatorial factors in the last expression, including the $\frac{1}{d_j}$ which is present in the orthonormality relations from the Peter-Weyl theorem (see appendix \ref{coherent_state_appendix} where we collect some definitions concerning representations of $\SU(2)$). Thus the Haar measure in terms of spinors is \emph{exactly} a normalized Gaussian measure.

\subsection*{$\SU(2)$-invariant observables and intertwiner space} \label{F-variables}
The spinorial variables are especially useful to extract the $\SU(2)$-gauge invariant content of the LQG Hilbert space on a given graph. Consider the following setup: Let $\gamma$ be a graph with $V$ vertices and $E$ edges. Each edge $e$ connects two vertices $v_I$ and $v_J$ and thus associates a spinor $\ket{z}$ to $v_I$ and a spinor  $\ket{\tz}$ to $v_J$ respectively. This means that an $N$-valent vertex carries $N$ spinors $\ket{z_1}, \dots \ket{z_N}$ which are associated to either beginning points of edges or final points of edges.\\
Concentrate on the subspace of spinors associated to one single $N$-valent vertex. Under $\SU(2)$-gauge transformations all the spinors living on that vertex transform as
\be
\ket{z_i} \stackrel{\SU(2)}{\mapsto} h_v\ket{z_i} \; \mbox{ for } h_v \in \SU(2) \mbox{ and } i = 1,\dots,N \; .\nn
\ee
Because they all transform in the same way it is easy to construct classical $\SU(2)$-invariant quantities as \cite{etera_spanish_laurent}
\be
E_{ij} := \braaket{z_i}{z_j}, \quad F_{ij} := \braaketrt{z_i}{z_j} \nn \; .
\ee
The $E_{ij}$ are symmetric $N\times N$ matrices over $\C$, however they mix the holomorphic and anti-holomorphic part of the spinors. The $F_{ij}$ are antisymmetric $N\times N$ matrices and holomorphic\footnote{For a detailed analysis of these $E$- and $F$-variables which parameterize the classical $\SU(2)$-invariant phase space at one single vertex see \cite{laurent_etera_un1, laurent_etera_un2, etera_spanish_laurent}. The $E$-variables form a $U(N)$-Poisson-sub-algebra, and the $F$-variables can be interpretated as generalized ladder operators. However, they are not completely independent from each other. They fulfill certain relation analogous to the Pl\"ucker relations of $\SU(2)$-recoupling theory.} They play a major role in understanding the $\SU(2)$-invariant Hilbert space of loop quantum gravity, as we will see in section \ref{su2-invariance}.\\
From the phase space point of view, the $\SU(2)$-invariance on each vertex can be understood as arising from a constraint at each vertex,
\be \label{su2_constraint}
\vec{\fc} := \sum\limits_{i=1}^N \vec{X}_i \; .
\ee
The action of this constraint on any spinor is simply given by
\be
\{  \vec{\fc},  \ket{z_i} \}  = \vec{\sigma} \ket{z_i} \; ,\nn
\ee
and thus finite gauge transformations generated by $\vec{\fc}$ are simply $\SU(2)$ transformations on each spinor separately,
\be
\ket{z_i} \stackrel{\SU(2)}{\mapsto} h_v \ket{z_i} \; ,\nn
\ee
where $h_v := e^{i\vec{p}\vec{\sigma}}$ is the finite group element at vertex $v$ for some choice of coordinate system $\vec{p}$ on the group.
From a geometric point of view, due to the identification of each spinor with a three-dimensional vector (\ref{zandX}), imposing $\vec{\fc} = 0$ just means demanding closure. \\
Consider the simplest possible situation, namely the one-loop graph (see figure \ref{oneloopgraph}): one single vertex connected by a closed loop. I.e. there are two spinors $\ket{z_1}$ and $\ket{z_2}$ living on that vertex which both behave under $\SU(2)$ transformations as $\ket{z_i} \stackrel{\SU(2)}{\mapsto} g_v \ket{z_i}$. A natural classical  gauge invariant quantity is the Wilson loop $W(g)$ around that loop, that is, just the trace over the group element $g(z_1, z_2)$ associated to the two spinors and given explicitly in (\ref{group_element}). This can easily be calculated:
\be \label{wilson_loop}
W(g) &  =  &\frac{\left( \braaketrt{z_2}{z_1}  +  \braakettr{z_2}{z_1}  \right) }{\norm{z_1} \norm{z_2}} = \frac{\left( -F_{12} + \overline{F}_{12}  \right) }{  \sqrt{E_{11} E_{22}}  } \; \nn \\
& = & -2 i \frac{ \Im(F_{12}) }{ \sqrt{E_{11} E_{22}} } \; .
\ee
More general holonomy functions (involving loops over an arbitrary number of vertices) were constructed in \cite{etera_spanish_laurent}, but the general idea is exactly the same. Depending on the exact configuration one ends up with an expression involving more or less complicated (sums of) products of $F$-operators living on all the different vertices through which the loop runs.\\

\section{$L^2(\SU(2),dg)$ in terms of spinors} \label{quantum_spinors}
Regarding the spinorial variables described in the last section as fundamental one is led to a different quantization: The most natural Hilbert space for these variables is the Bargmann space of holomorphic square integrable complex functions, on which the complex numbers $z$ and $\bz$ act as multiplication and derivation operators respectively. Taking the $\U(1)$-invariance on each edge into account one is led to a space $\cH_e^{\rm spin} := L_{\rm hol}^2(\C^4, d\mu)/ \U(1)$ which we will describe in the following. However, as working with spinorial variables amounts to choosing a different polarization of the classical phase space it is a priori unclear (see for example \cite{woodhousebook}) whether this space is unitarily equivalent to the standard LQG Hilbert space associated to a single edge $\cH_e := L^2(\SU(2), dg)$. We answer this question in the affirmative by explicitly constructing a unitary map between the two.\\
In this section we consider the Hilbert spaces associated to one single edge only and explore in detail the properties of our unitary map. In the next section we generalize this map to an arbitrary graph, show that our construction is cylindrically consistent (hence, the continuum limit can be performed) and analyze in detail how to take $\SU(2)$-invariance at the vertices into account.

\subsection{Bargmann space}
\subsubsection*{Bargmann space over one complex variable}
Consider the Hilbert space of holomorphic, square-integrable functions \cite {bargmann1, bargmann_reps} over $\C$: $\cF := L^2_{\rm hol}(\C, d\mu)$, where the measure\footnote{There actually is some freedom in the choice of measure which can be used to absorb some combinatorial factors in the basis states, see appendix \ref{factor_dj_appendix}.} is a Gaussian one given by
\be
d\mu(z) := \frac{1}{\pi}e^{-|z|^2}dz \, .\nn
\ee
$\mu$ is normalized, $\int\limits_\C d\mu(z) = 1 $, invariant under translations $\mu(z+a) = \mu(z) \; \forall a \in \C$, and under inversion, $\mu(-z) = \mu(z)$.
Elements of $\cF$ are holomorphic functions, i.e. they can be expanded in powers of $z$ (and do not depend on the complex conjugate $\bar{z}$):
\be
\forall f \in \cF: \; f(z):= \sum\limits_{n=0}^\infty \alpha_n z^n \nn
\ee
with complex coefficients $\alpha_n$. The inner product between two functions $f,f' \in \cF$ is thus given by
\be
 \int\limits_\C d\mu(z) \overline{f(z)} f'(z) \; ,\nn
\ee
and an orthonormal basis of $\cF$ is provided by
\be
e_n(z) := \frac{1}{\sqrt{n!}}z^n\nn
\ee
as can easily be checked.\\
The natural symplectic structure of $\C$,
\be \label{sympl_C}
\{ \bz, z  \} = i\; ,
\ee
is represented on $\cF$ in an intuitive way. Assume $a, a^\dagger \in \cL(\cF)$ acting as
\be
(a f)(z) := zf(z), \qquad  (a^\dagger f)(z) := \partial_z f(z), \qquad \forall f \in \cF \; ,\nn
\ee
then they fulfill the algebra\footnote{Here we follow the slightly non-standard conventions of \cite{bargmann1, bargmann_reps} in the definition of $a$ and $a^\dagger$.}
\be
[a^\dagger, a]= 1 \; ,\nn
\ee
which clearly is a representation of (\ref{sympl_C}).
The delta-distribution on $\cF$ is given by
\be
\delta_w(z) := e^{z\bar{w}} \text{ \rm for } w \in \C\; ,\nn
\ee
in the sense that for every function $f\in \cF$ we have
\be
\int d\mu(z) \overline{\delta_w(z)}  f(z)  = f(w).\nn
\ee
$\delta_w$ is a proper element of $\cF$, because it is obviously holomorphic, and also square-integrable
\be
\int d\mu(z) \overline{\delta_w(z)} \delta_{w'}(z) = e^{w\bar{w'}} \; .\nn
\ee
The completeness relation in the basis $e_n(z)$ is obviously
\be
\sum\limits_{n\in\N}\overline{e_n(z')}e_n(z) = e^{z\overline{z}'} = \delta_{z'}(z)\nn
\ee

\subsubsection*{Bargmann space over 2 complex variables and spinor states}
The Bargmann-space can easily be generalized to $n$ copies of the complex plane, where it takes the form $\cF_n := L^2_{\rm hol}(\C^n, d\mu(z))$ with an appropriate normalized Gaussian measure on $\C^n$. Let us analyze the space $\cF_2$ a bit further: This space carries a natural representation of $\SU(2)$, by push-forward from the fundamental representation of $\SU(2)$ on $\C^2$. An orthonormal basis of this space is given by the polynomials $\frac{1}{\sqrt{a! b!}}(z^0)^a (z^1)^b$ with $a,b \in \N$. Or alternatively, in the more familiar notation from $\SU(2)$ representation theory, one can define an orthonormal basis as
\be \label{basis_jm}
e^j_m(z) := \frac{1}{\sqrt{(j+m)!(j-m)!}} (z^0)^{j+m} (z^1)^{j-m}\; ,
\ee
with $j := \frac{1}{2}(a+b)$, $m := \frac{1}{2}(a - b)$ and thus $j \in \frac{1}{2}\N$ and $-j<m< +j$. This is the analogue of the standard magnetic number basis used in $\SU(2)$-representation theory (see also appendix \ref{coherent_state_appendix} where we collect some useful definitions and formulae for different bases of $\SU(2)$). The $e^j_m(z)$ are easily seen to be orthogonal with respect to the inner product on $\cF_2$.\\
Note that the space $\cF_2$ decomposes into $d_j := (2j+1)$-dimensional subspaces $\cD^j_2$ of homogeneous polynomials of degree $2j$ as
\be
\cF_2 = \mathop{\oplus}\limits_{j\in \N/2}\cD_2^j \; ,\nn
\ee
where each space $\cD_2^j$ is spanned by the basis elements $e^j_m(z)$ with fixed $j$. In particular, the delta-distribution on each subspace $\cD^j_2$ is given by
\be
\delta^j_{\ket{\omega}}\ket{z} := \sum\limits_{m = -j}^{j} \frac{(\bar{\omega}^0 z^0)^{j+m}(\bar{\omega}^1 z^1)^{j-m}}{(j+m)!(j-m)!} = \frac{1}{(2j)!}\braaket{\omega}{z}^{2j}\; ,\nn
\ee
in the sense that for each $f^j \in \cD^j_2$ we have
\be
\int d\mu(z) \overline{\delta^j_{\ket{\omega}}\ket{z}} f^j(z) = f^j(\omega) \; .\nn
\ee
The harmonic oscillator operators act on the basis (\ref{basis_jm}) as
\be \label{osc_rep}
(a^0)^\dagger e^j_m(z) = z^0 e^j_m(z)  & = & \sqrt{j+m+1}e^{j+\frac{1}{2}}_{m+\frac{1}{2} }(z) \nn \\
(a^1)^\dagger e^{j}_{m}(z) = z^1 e^j_m(z)  & = & \sqrt{j-m+1}e^{j+\frac{1}{2}}_{m-\frac{1}{2} }(z) \nn \\
(a^0) e^j_m(z) = \partial_{z^0} e^j_m(z)  & = & \sqrt{j+m}e^{j-\frac{1}{2}}_{m-\frac{1}{2} }(z) \nn \\
(a^1) e^j_m(z) = \partial_{z^1} e^j_m(z)  & = & \sqrt{j-m}e^{j-\frac{1}{2}}_{m+\frac{1}{2} }(z) \; .
\ee
Now let's have a look at the representation of $\SU(2)$ on $\cF_2$:  This can easily be defined by push-forward from the fundamental representation of $\SU(2)$ on $\C^2$. For each  group element $g \in \SU(2)$ define
\be
D(g): \cF_2 \rightarrow \cF_2; \quad f(z) \mapsto (D(g)f)(z) := f(\,^tg z) \; ,\nn
\ee
where $\;^tg $ is the transpose of $g$. Unitarity of this representation holds because the measure $d\mu(z)$ is clearly invariant under $\SU(2)$-transformations. Furthermore, it leaves the subspaces $\cD_2^j$ invariant.

\subsection*{U(1)-gauge invariant states and the spinorial edge-Hilbert space} \label{u1_inv_spinors}
Consider one single edge with spinors $\ket{z}$ and $\ket{\tilde{z}}$ living on the beginning and the end of the edge respectively. The corresponding Hilbert space for one edge is then given by $\cF_2 \otimes \cF_2 := L^2_{\rm hol}(\C^2, d\mu) \otimes L^2_{\rm hol}(\C^2, d\mu)$. However, from the loop quantum gravity perspective this space is too big, because the matching condition (\ref{u1_constraint}) is not taken into account yet.  Thus, the spinorial Hilbert space associated to one single edge should be given by
\be
\cH_e^{\rm spin} :=\cF_2\times \cF_2/\U(1) \; .\nn
\ee
What do the elements of this space look like? Let's start with an arbitrary (normalized, orthogonal) basis-element of $\cF_2 \otimes \cF_2$
\be
e^{j_1}_{m}(z) \otimes e^{j_2}_{n}(\tilde{z}) := \frac{(z^0)^{j_1 +  m}  (z^1)^{j_1 - m}  (\tilde{z}^0)^{j_2 + n}  (\tilde{z}^1)^{j_2 - n}  }{\sqrt{ (j_1 + m)!  (j_1 - m)!  (j_2  + n )!  (j_2 - n)!       }}  \; .\nn
\ee
They behave under the action of $\U(1)$ as
\be
e^{j_1}_{m}(z) \otimes e^{j_2}_{n} \stackrel{\U(1)}{\rightarrow} e^{i[2(j_1 - j_2)\phi]}e^{j_1}_{m}(z) \otimes e^{j_2}_{n} \; .\nn
\ee
Applying group averaging to construct gauge invariant states one sees that these do only exists for $j_1 = j_2$. All states with $j_1 \neq j_2$ get identically mapped to zero. For $j_1 = j_2$ the above states are already invariant and we define
\be
\cP^{j}_{mn }(z, \tilde{z}) :=  e^{j}_{m}(z) \otimes e^{j}_{n}(\tilde{z})  \; .\nn
\ee
The $\cP^{j}_{mn}$ form an orthonormal basis of $\cH^{\rm spin}_e$ which can in turn be defined as the completion of the linear span of $\cP^{j}_{mn}$. Every function $f \in \cH^{\rm spin}_e$ can be decomposed as
\be
f(z,\tilde{z}) = \sum\limits_{ j \in \N/2} \sum\limits_{m,n=-j}^{j} \hf^j_{mn}\cP^{j}_{mn}(z,\tilde{z}) \, ,\nn
\ee
where $\hf^j_{mn} := \int d\mu(z) d\mu(\tz) \overline{\cP^{j}_{mn}(z, \tz)} f(z, \tz)$ are the coefficients of $f$ in that basis. Orthonormality and completeness in $\cH^{\rm spin}_e$ read as
\be
 \int d\mu(z) d\mu(\tz)\overline{\cP^{j}_{mn}(z,\tz)} \cP^{j'}_{m'n'}(z,\tz) & = & \delta_{jj'}\delta_{mm'}\delta_{nn'} \nn \; , \\
 \sum\limits_{ j \in \N/2} \sum\limits_{m,n=-j}^{j} \cP^{j}_{mn}(z_1, \tz_1)\overline{\cP^{j}_{mn}(z_2, \tz_2)} & = & \delta^{\U(1)}_{(\ket{z_2}, \ket{\tz_2})}(\ket{z_1}, \ket{\tz_1 })\, ,\nn
\ee
where the delta-distribution in $\cH^{\rm spin}_e$ is given by
\be
\delta^{\U(1)}_{(z_2, \tz_2)}(z_1, \tz_1)
 & := & \sum\limits_{j \in \N/2} \frac{1}{[(2j)!]^2}\braaket{z_2}{z_1}^{2j}\braaket{\tz_2}{\tz_1}^{2j} \nn \\
& = & I_0(2\braaket{z_2}{z_1}\braaket{\tz_2}{\tz_1}) \; ,\nn
\ee
and $I_0(x)$ is the zeroth modified Bessel function of first kind.

\subsubsection*{Spinor coherent states}
The states $\cP^j_{mn}(z, \tz) \in \cH_e^{\rm spin}$ are the analogues of the magnetic number basis states used in $\SU(2)$-representation theory. However, for some computations it will also be useful to work with the spinorial analogue of the $\SU(2)$-coherent state basis (see appendix \ref{coherent_state_appendix} for definitions). Thus, starting from the orthonormal basis $\cP^j_{mn}(z,\tz)$ and using two additional spinors $\ket{\omega}, \ket{\tilde{\omega}} \in \C^2$ we define
\be
\cP^{j}_{\tilde{\omega} \omega}(z, \tz) := \sum\limits_{m,n= -j}^{+j} \frac{(2j)!(\omega^0)^{j+m}(\omega^1)^{j-m}(\bar{\tilde{\omega}})^{j+n}(\bar{\tilde{\omega}})^{j-n}}{\sqrt{(j+m)!(j-m)!(j+n)!(j-n)!}} \cP^j_{mn}(z, \tilde{z}) \, .\nn
\ee
The sums can performed and we obtain simply
\be
\cP^j_{\tilde{\omega} \omega} = \frac{1}{(2j)!}\braaket{\tilde{\omega}}{z}^{2j}\brar{\tz}\epsilon \ket{\omega}^{2j} \, , \nn
\ee
where $\epsilon$ is again the antisymmetric tensor introduced in section \ref{spinor_variables}.
The completeness relations on $\cH_e^{\rm spin}$ in terms of these spinor coherent states can be derived as
\be \label{completeness_coherent_spinors}
\int d\mu(z) \int d\mu(\tz) \overline{\cP^j_{\omega \tilde{\omega}}(z,\tz)} \cP^{k}_{\alpha \tilde{\alpha}}(z,\tz) & = & \frac{\delta^{jk}}{d_j}\braaket{\alpha}{\omega}^{2j}\braaket{\tilde{\omega}}{\alpha}^{2j} \, ,\nn \\
\sum\limits_j \int d\mu(\omega) d\mu(\tilde{\omega}) \frac{d_j}{(2j)!} \overline{\cP^j_{\omega \tilde{\omega}}(z_1, \tz_1)} \cP^j_{\omega \tilde{\omega}}(z_2, \tz_2) & = & I_0(2\braaket{z_1}{z_2} \braaket{\tz_1}{\tz_2}) \, .
\ee

\subsection{A unitary map} \label{sec:unitary_map}
The situation obtained so far is the following: We started with a classical phase space $T^*\SU(2)$ and two different polarizations: the first one in terms of group elements and Lie-algebra elements $(g,X)$ and the second one in terms of spinors $(\ket{z}, \ket{\tz})$. A quantization based on the first polarization leads to the standard edge Hilbert space $\cH_e = L^2(\SU(2), dg)$, a quantization based on the second to the spinor Hilbert space $\cH_e^{\rm spin} = \cF_2 \otimes \cF_2 / \U(1)$. The question is whether these two spaces carry the same physical information, i.e. whether they are unitarily equivalent. In general, two quantizations of the same phase space based on different polarizations are unitarily inequivalent \cite{woodhousebook}. However, in the present case we can explicitly construct such a unitary map without much effort\footnote{See also appendix \ref{segal_bargmann_appendix} where we review the Segal-Bargmann transform on the real line which is very similar to our construction.}.\\
Using the basis $\cP^{j}_{mn}$ of $\cH^{\rm spin}_e$ and the standard Peter-Weyl decomposition of $\cH_e$ (\ref{PW}) it is obvious that there exists a unitary map between the two spaces.
Because they fulfill the same orthogonality and completeness relations (see appendix \ref{coherent_state_appendix}), we can define a unitary map on the basis states between both spaces\footnote{The factor $\frac{1}{\sqrt{d_j}}$ has to be included to get the same orthonormality as in the group case. Alternatively it could be absorbed in the measure on $\C^2 \times \C^2$. See appendix \ref{factor_dj_appendix} for further discussions and an explanation of the combinatorial origin of this factor.},
\be
\cT: && \cH_e \rightarrow \cH^{\rm spin}_e; \\
&& D^j_{mn}(g) \mapsto (\cT D^j_{mn})(z, \tilde{z}) := \frac{1}{ \sqrt{d_j}}\cP^{j}_{mn}(z, \tilde{z}) \; .\nn
\ee
For a general function $f \in \cH_e$ with Fourier decomposition given by (\ref{PW}) the map is then simply
\be
\cT: && \cH_e \rightarrow \cH^{\rm spin}_e; \nn \\
&& f \mapsto  (\cT f)(z,\tilde{z})  :=   \sum\limits_{j \in \N/2} \sum\limits_{m,n=-j}^{j} \hf^j_{mn} \cP^{j}_{mn}(z, \tilde{z}) \; . \nn
\ee
This establishes unitary equivalence between the Hilbert spaces $\cH_e = L^2(\SU(2), dg)$ and $\cH^{\rm spin}_e = \cF_2 \times \cF_2 / \U(1)$.
In terms of an integral kernel this map can be written as
\be
(\cT f)(z, \tz) := \int dg \cK_g(z, \tz) f(g) \; ,\nn
\ee
where the kernel $\cK_g(z, \tilde{z})$ is defined as
\be
\cK_g(z, \tz) :=  \sum\limits_{j \in \N/2} \sum\limits_{m,n=-j}^{j}\sqrt{d_j} \overline{D^j_{mn}(g)} \cP^{j}_{mn}(z, \tilde{z}) \; .\nn
\ee
Using coherent states instead of the basis $\cP^j_{mn}$ this kernel can be written in a more compact form. As the $\SU(2)$-coherent states $D^j_{\omega \tilde{\omega}}(g)$ are mapped onto the spinor coherent states $\cP^j_{\omega \tilde{\omega}}(z, \tz)$ (up to the factor of $\frac{1}{\sqrt{d_j}}$ again) the kernel can equally well be written as
\be \label{kernel}
\cK_g(z, \tz) & = & \sum\limits_j \int d\mu(\omega) d\mu(\tilde{\omega}) \frac{\sqrt{d_j}}{[(2j)!]^2} \overline{D^j_{\omega \tilde{\omega}}(g)} \cP^j_{\omega \tilde{\omega}}(z, \tz) \nn \\
& = & \sum\limits_j \int d\mu(\omega) d\mu(\tilde{\omega}) \frac{\sqrt{d_j}}{[(2j)!]^3}\braaaket{\tilde{\omega}}{g^{-1}}{\omega}^{2j} \brar{\tz}\epsilon\ket{\tilde{\omega}}^{2j}\braaket{\omega}{z}^{2j} \nn \\
& = & \sum\limits_{k \in \N} \frac{\sqrt{k+1}}{k!} \brar{\tz}\epsilon g^{-1} \ket{z}^k \, ,
\ee
which is almost an exponential up to the factor of $\sqrt{k+1}$. In appendix \ref{factor_dj_appendix} we explain how this factor can be absorbed in a change of measure to really turn the kernel into an exponential function. However, for the moment we will stick to this definition.\\
Writing the unitary map in terms of coherent states allows for a very intuitive interpretation: note that the group element $g(z, \tz)$ splits into a holomorphic and an antiholomorphic part when written in terms of spinors as
\be
g(z, \tz) = \frac{\ket{z} \brar{\tz} - \ketr{z}\bra{\tz}}{\norm{z} \norm{\tz}}\, . \nn
\ee
It is easy to see that the map $\cT$ essentially (up to a an inclusion of $\epsilon$) restricts the representation matrices of $\SU(2)$, written in terms of spinors, to their holomorphic part:
\be \label{restriction_hol}
 D^j_{\omega \tilde{\omega}}(g) = \left( \bra{\omega} \frac{\ket{z}\brar{\tz} - \ketr{z}\bra{\tz} }{\sqrt{\braaket{z}{z} \braaket{\tz}{\tz}}}  \ket{\tilde{\omega}}  \right)^{2j} \stackrel{\cT}{\mapsto} \frac{1}{(2j)!\sqrt{d_j}}\braaket{\omega}{z}^{2j}\brar{\tz} \epsilon \ket{\tilde{\omega}}^{2j} \, .
\ee
To summarize, the unitary map $\cT: L^2(\SU(2)) \rightarrow \cF_2 \otimes \cF_2/\U(1)$ we found can be written in three different forms (in terms of the magnetic number basis, in terms of the coherent state basis, or in terms of an integral kernel $\cK_g(z, \tz)$) and it depends on the kind of computation performed which one is the most useful:
\be \label{unitary_transform}
 D^j_{mn}(g) \qquad & \stackrel{\cT}{\mapsto} &  \qquad \frac{1}{\sqrt{d_j}} \cP^{j}_{mn}(z, \tz), \nn \\
 \braaaket{\omega}{g}{\tilde{\omega}}^{2j} \qquad & \stackrel{\cT}{\mapsto} & \qquad \frac{1}{\sqrt{d_j}(2j)!} \braaket{\omega}{z}^{2j} \brar{\tz} \epsilon\ket{\tilde{\omega}}^{2j} \nn \\
 f(g) \qquad & \stackrel{\cT}{\mapsto} & \qquad \int dg \cK_g(z,\tz) f(g) \, .
\ee
Note that the factor of $\frac{1}{\sqrt{d_j}}$ is important to ensure that $\cT$ is unitary: this way the scalar product between two spinor basis states is exactly dual to the scalar product between group representation matrices in the Peter-Weyl theorem.

\subsection{Group multiplication and convolution property} \label{convolution_sec}
In terms of $\SU(2)$-coherent states the group multiplication property (\ref{group_multiplication}) for representation matrices can be written as
\be
\int d\mu(\alpha) D^j_{\omega \alpha}(g_1) D^j_{\alpha \tilde{\omega}}(g_2) & = & (2j)!D^j_{\omega \tilde{\omega}}(g_1 g_2) \, .\nn
\ee
It turns out that on $\cH_e^{\rm spin}$ such a property does not hold anymore. Instead we obtain
\be \label{modified_group_law}
\int d\mu(\alpha) \cP^j_{\omega \alpha}(z_1, \tz_1) \cP^j_{\alpha \tilde{\omega}}(z_2, \tz_2) = \cP^j_{\omega \tilde{\omega}}(z_1, \tz_2) \brar{\tz_1}\epsilon \ket{z_2}^{2j}
\ee
In abstract terms, the unitary map $\cT$ is not a group homomorphism from the space of representation matrices onto the spinor space\footnote{See also the article \cite{flux_representation} where a similar issue was encountered when rewriting the kinematical Hilbert space of loop quantum gravity in terms of non-commutative flux variables.}. This can again be understood by looking at the holomorphic-antiholomorphic splitting of the group element. $\cT$ restricts group representation matrices to their holomorphic part, and the product of two such holomorphic projections necessarily still is holomorphic. However, the product fails to be in the image of $\cT$, i.e. cannot be written as the holomorphic projection of a third group element.\\
An interesting question is whether the kernel of the unitary transform $\cK_g(z,\tz)$ fulfills some ``convolution property'':
\be
\cK_{gh}(z, \tz) & = & \sum\limits_j \frac{\sqrt{d_j}}{(2j)!} \brar{\tz}\epsilon h^{-1}g^{-1}\ket{z}^{2j} \nn \\
& = & \int d\mu(\omega) \sum\limits_j \sum\limits_k \frac{\sqrt{d_j}}{(2j)!(2k)!} \brar{\tz} \epsilon h^{-1} \ket{\omega}^{2j} \brar{\bar{\omega}} \epsilon g^{-1} \ket{z}^{2k} \nn \\
& = & \int d\mu(\omega)  \cK_g(z, \bar{\omega}) e^{\brar{\tz} \epsilon h^{-1} \ket{\omega}} \nn 
\ee
where in the second line we used the identity $\int d\mu(\omega) \frac{1}{(2j)!} \braaket{z}{\omega}^{2j}\braaket{\omega}{\tz}^{2k} = \delta^{jk}\braaket{z}{\tz}^{2j}$ and that $\bra{\bar{z}} = \brar{z}\epsilon$. The missing ingredient to make this convolution \emph{exactly} dual to the group multiplication is again a factor of $\sqrt{d_j}$. One could replace the kernel with a different one by ``weighting'' it with a factor of $\frac{1}{\sqrt{d_j}}$ according to
$\cK_g(z, \tz) \rightarrow \cK^+_g(z, \tz) := e^{\brar{\tz} \epsilon g^{-1} \ket{z}}$. This makes the kernel exactly dual to the group multiplication as
\be
\cK^+_{gh}(z, \tz) = \int d\mu(\omega) \cK^+_g(z, \bar{\omega}) \cK^+_h(\omega, \tz) =: (\cK_g \circ \cK_h)(z, \tz) \, . 
\ee
However, in order to accomodate for such a change in the kernel, one would also have to change the measure on $\C^2 \times \C^2$ in order to conserve the orthonormality-properties of the unitary transform. For the purpose of this paper we will thus stick to the definition (\ref{kernel}). We refer the reader to appendix \ref{factor_dj_appendix} for a more detailed discussion of that issue.

\subsection{Operators on the spinor space}
Regarding the spinorial variables as fundamental changes the standard picture a bit: The most natural operators acting on $\cH^{\rm spin}_e$ are simply the ladder operators $(\hat{a}^\dagger, \hat{\tilde{a}}^\dagger, \hat{a}, \hat{\tilde{a}})$ associated to the spinors $(z, \tz, \bar{z}, \bar{\tz})$ as defined in (\ref{osc_rep}). However, these operators are neither $\U(1)$- nor $\SU(2)$-invariant. When demanding $\U(1)$-invariance, one can define the associated holonomy- and flux-operators simply by pull-back from the loop quantum gravity Hilbert space with $\cT$: let $\psi \in \cH_e$ and $\psi^{\rm spin} := (\cT \psi)(z, \tz)$. Then the action of fluxes and cylindrical functions (of which holonomies are a special case) on the spinor states is given by
\be \label{definition_operators}
[\hat{X}^i\psi^{\rm spin}](z, \tz)  :=   [ \cT (\hat{X}^i \psi)](z, \tz), \qquad
[\hat{f} \psi^{\rm spin}](z, \tz)    :=  [ \cT (\hat{f} \psi)](z, \tz) \; .
\ee
However, written in in terms of ladder-operators, these will now be composite operators consisting of creation- as well as annihilation operators. For the fluxes these can be given easily:
\be
\hat{X}^1 = (\hat{a}^0)^\dagger \hat{a}^1 + (\hat{a}^1)^\dagger \hat{a}^0 , \; \hat{X}^2 = (\hat{a}^0)^\dagger \hat{a}^1 - (\hat{a}^1)^\dagger \hat{a}^0 , \; \hat{X}^3 = (\hat{a}^0)^\dagger\hat{a}^0 - (\hat{a}^1)^\dagger \hat{a}^1 \; .
\ee
For the holonomy-operators and all other functions $\hat{f}(g)$, which just act by multiplication on $L^2(\SU(2))$, the situation is more complicated. In principle the strategy is the following: one should decompose
\be
f\psi(g) := \sum\limits_{j,m,n}\sqrt{d_j} (f\psi)^j_{mn} D^j_{mn}(g) \, ,
\ee
where the coefficients $(f\psi)^j_{mn}$ are rather complicated expressions that can be obtained from $\SU(2)$-recoupling theory. As the unitary map $\cT$ maps $D^j_{mn}(g) \mapsto \frac{1}{\sqrt{d_j}}\cP^j_{mn}(z,\tz)$ the unitary transform of $(f\psi)(g)$ will be just
\be
(f\psi)(z,\tz) := \sum\limits_{j,m,n} (f\psi)^j_{mn}\cP^j_{mn}(z,\tz) \, ,
\ee
and thus the action of cylindrical functions (of which holonomies are a special case) is given implicitly by
\be
[\hat{f} \psi^{\rm spin}](z, \tz) := \sum\limits_{j,m,n}(f\psi)^j_{mn} \cP^j_{mn}(z, \tz) \, .
\ee
However, as it stand this expression is not of much use as the explicit form of the above operator (i.e. the exact form of the coefficients $(f\psi)^j_{mn}$) can only be obtained through recoupling theory.\\
An alternative route, which in many cases might be less cumbersome, is the following: Simply write
\be
f(g) = f(g(z,\tz)) \, ,
\ee
using the description of $g$ in terms of spinors. Then one obtains the corresponding quantum operator on the spinor Hilbert space $\cH^{\rm spin}_e$ by simply replacing spinors with the associated creation- and annihilation operators as $(z, \tz, \bar{z}, \bar{\tz}) \rightarrow (\hat{a}^\dagger, \hat{\tilde{a}}^\dagger, \hat{a}, \hat{\tilde{a}})$. However, depending on the complexity of $f$ in terms of spinors, one will have to carefully choose a meaningful operator ordering in order to make this prescription compatible with (\ref{definition_operators}). A good starting point for such an analysis seem Wilson loop operators, analogous to the classical function described in (\ref{wilson_loop}). We leave this issue open for future research.\\
A different class of operators, when considering the spinor Hilbert space $\cH^{\rm spin}_\gamma$ associated to a larger graph are the quantum analogues of the $E$- and $F$-variables described in section \ref{F-variables}. They are $\SU(2)$-invariant observables living at each vertex of the graph $\gamma$ and provide a useful tool to understand the intertwiner spaces associated to that graph (see \cite{laurent_etera_un1, laurent_etera_un2, etera_spanish_un1, etera_spanish_un2, etera_spanish_laurent}).

\subsection{Examples}
To illustrate how the unitary map $\cT$ works in detail, we want to give some examples. In this paragraph we restrict ourselves to one single copy of the gauge group (i.e. the graph consists of one single edge only) and compute the transform of (i) characters on the group, (ii) the delta distribution on the group, and (iii) the heat kernel.
In section \ref{su2-invariance} we will have a look at more general graphs and focus on the issue of $\SU(2)$-invariance, where the spinors turn out be especially useful.

\begin{itemize}
\item The characters $\chi^j(g)$ of $\SU(2)$ can be written either in the magnetic number basis or in the coherent state basis as
\be
\chi^j(g) & = & \sum\limits_{m} D^j_{mm}(g) \nn \\
& = & \frac{1}{(2j)!}\int d\mu(\omega) \braaaket{\omega}{g}{\omega}^{2j} \, .\nn
\ee
Using either form of $\cT$ (in terms of the magnetic number basis, in terms of the coherent state basis or in terms of the integral kernel) it is easy to verify that the characters are mapped onto the $\delta^j$-distributions (up to a factor of the square root of the dimension of the representation space):
\be
\chi^j(g) \;  \mapsto \; (\cT \chi^j)(z,\tz) =\frac{1}{\sqrt{d_j}}\frac{1}{(2j)!} \brar{\tz}\epsilon \ket{z}^{2j} = \frac{1}{\sqrt{d_j}}\delta^j_{\ket{\bar{\tz}}}\ket{z}\nn
\ee

\item The delta-distribution on the group can simply be written as
\be
\delta_h(g) = \sum\limits_{j,m,n}d_j\overline{D^j_{mn}(h)}D^j_{mn}(g) = \sum\limits_{j} d_j\chi^j(h^{-1}g) \, .\nn
\ee
Using the group multiplication property \ref{group_mult_coherent} and considering $\delta_h(g)$ as a function of $g$ for fixed $h$ it is not difficult to verify that
\be
\delta_h(g) \mapsto (\cT_g \delta_h(g))(z, \tz) & = & \sum\limits_{j} \frac{\sqrt{d_j}}{(2j)!} \brar{\tz} \epsilon {h^{-1}} \ket{z}^{2j} \nn \\
& = & \cK_h(z, \tz) \, ,\nn
\ee
where $\cK_h(z,\tz)$ is the kernel of the transform evaluated at $h$.\\
Applying the adjoint transform\footnote{Here the adjoint of the transform is needed because the group element $h$ only appears in the adjoint basis $\overline{D^j_{mn}(g)}$ in the delta-distribution. The adjoint transform is constructed such that it is compatible with taking the complex conjugate. In terms of an integral kernel this can be written as $\cK^\dagger_h(\omega, \tilde{\omega}) = \sum\limits_j \frac{\sqrt{d_j}}{(2j)!}\bra{\omega} h \epsilon^{-1} \ketr{\tilde{\omega}}^{2j}$.} $\cT^\dagger$ to the second group variable $h$ one obtains
\be
(\cT^\dagger_h  \cT_g  \delta_h(g))(z,\tz, \omega, \tilde{\omega}) & = & \sum\limits_j \frac{1}{[(2j)!]^2}\braaket{\omega}{z}^{2j}\braaket{\tilde{\omega}}{\tilde{z}}^{2j} \nn \\
& = & I_0(2\braaket{\omega}{z}\braaket{\tilde{\omega}}{\tz}),\nn
\ee
where $I_0$ is the zeroth modified Bessel function of first kind that plays the role of the delta-distribution on $\cH_e^{\rm spin}$ as explained in section \ref{u1_inv_spinors}.

\item The heat kernel $H_t(h, g)$ is defined as the solution to the heat equation
\be
(\partial_t - \Delta)H_t(h,g) = 0, \qquad H_0(h,g) = \delta_h(g) \, .\nn
\ee
Explicitly, on $\SU(2)$, it can be written in terms of characters as
\be
H_t(g,h) = \sum\limits_j d_j e^{tj(j+1)} \chi^j(h^{-1}g) \, .\nn
\ee
Exactly analogous to the above example one obtains for its unitary transform
\be
(\cT^\dagger_h \cT_g  H_t(g,h) )(z,\tz, \omega, \tilde{\omega}) = \sum\limits_j \frac{d_j}{[(2j)!]^2} e^{tj(j+1)}\braaket{\omega}{z}^{2j}\braaket{\tilde{\omega}}{\tz}^{2j} \, .\nn
\ee
The heat kernel of $\SU(2)$ plays a central role in the construction of Hall's coherent states \cite{hall} which have been used in the loop quantum gravity context due to their semi-classical behavior. It will be interesting to further analyze their spinor-analogous and their geometric meaning. We leave this point open for future research.
\end{itemize}

\section{Quantum gravity with spinors: arbitrary graphs and continuum} \label{sec:quantum_arbitrary}

In the last section we constructed a unitary map $\cT$ and showed that the Hilbert space associated to a single edge of a graph in loop quantum gravity can be mapped onto a spinor space $\cH^{\rm spin}_e := \cF_2 \times \cF_2/ U(1)$ in a unitary way.\\
Now let us move to a more complicated setting and consider an arbitrary graph $\gamma$ with $E$ edges (for an example see figure \ref{graph}). As long as $\SU(2)$-invariance at the vertices is not taken into account the Hilbert space associated to this graph factorizes\footnote{In principle this is also the case after taking $\SU(2)$-invariance into account as the gauge invariant Hilbert space is a true subspace of the gauge variant one. However, only certain combinations of functions are allowed, and these typically couple elements of all edge-Hilbert spaces linked to a vertex in a non-trivial way. } into $\cH_\gamma = \bigotimes\limits_e \cH_e$. Thus the unitary map can simply be generalized to an arbitrary graph $\gamma$ by letting it act on each edge-Hilbert space $\cH_e$ separately. Most conveniently this map can be written in terms of an integral kernel which is just a product of edge contributions. Given an arbitrary graph $\gamma$ let us associate to it a Hilbert space
\be
\cH_\gamma^{\rm spin} := \bigotimes\limits_e \cH_e^{\rm spin} \, .\nn
\ee
Then there is a unitary map between the loop quantum gravity Hilbert space $\cH_\gamma$ and the spinor Hilbert space as
\be
\cT_\gamma: \cH_\gamma & \rightarrow & \cH_\gamma^{\rm spin}; \nn \\
    f(g_1, \dots, g_E) & \mapsto & (\cT_\gamma f)(z_1, \tz_1, \dots ,z_e, \tz_E) := \int dg_1 \dots dg_E \cK^\gamma_{(g_1, \dots, g_E)}(z_1, \tz_1, \dots, z_E, \tz_E) f(g_1, \dots, g_E) \, , \nn
\ee
where the kernel is given by a product of edge contributions
\be
\cK^\gamma_{(g_1, \dots, g_E)}(z_1, \tz_1, \dots, z_E, \tz_E) := \prod\limits_{e} \cK_{g_e}(z_e, \tz_e) \, ,\nn
\ee
and the indices on the group elements and spinors label the edges in this case.\\
Using the results from the last section it is easy to see that, for each graph $\gamma$ separately, $\cH_\gamma$ and $\cH_\gamma^{\rm spin}$ are unitarily equivalent. That means, as long as keeping the graph fixed the description using spinors is physically equivalent to the description using group variables. That this is the case in the classical setting (i.e. equivalence of phase spaces) was shown already in \cite{laurentsimone}. Here we extended these results to the quantum theory.

\subsection{Cylindrical consistency and the continuum spinor space} \label{cyl_cons}
Having established unitary equivalence for each graph separately, it remains to analyze whether the maps $\cT_\gamma$ respect the conditions of cylindrical consistency, and thus, whether loop quantum gravity's continuum Hilbert space\footnote{See appendix \ref{lqg}, where we briefly recall the construction of that space. Heuristically speaking the conditions of cylindrical consistency assure that ``nothing depends on the graph''. This distinguishes the framework of loop quantum gravity from lattice approaches to quantum gravity and is the reason why the theory, despite being defined on graphs, can account for an infinite number of degrees of freedom. At the level of Hilbert spaces there exist certain isometric embedding $^*p_{\gamma \gamma'}: \cH_\gamma \rightarrow \cH_{\gamma'} \; \forall \gamma' \geq \gamma$. These are used to define an equivalence relation that relates states living on \emph{different} graphs. These equivalence classes, not individual states, are elements of the continuum theory Hilbert space $\cH$.} $\cH := \overline{\cup_\gamma \cH_\gamma / \sim}$ has a spinorial continuum analogue. It turns out that the answer is in the affirmative:\\
Whether the unitary equivalence can be lifted from the level of individual graphs to the full continuum Hilbert space $\cH$ depends on whether there exist suitable isometric embeddings $^*p^{\rm spin}_{\gamma \gamma'}$ between spinor Hilbert spaces $\cH^{\rm spin}_\gamma$ and $\cH^{\rm spin}_{\gamma'}$ defined on different graphs $\gamma \neq \gamma'$. But, having a unitary map $\cT_\gamma: \cH_\gamma \rightarrow \cH^{\rm spin}_\gamma$ for each graph $\gamma$, it is immediate to see that these can be \emph{defined} by demanding the following diagram to commute:
\be \label{diagram}
\xymatrix{
  \cH_\gamma \ar[r]^{\cT_{\gamma}} \ar[d]_{\,^*p_{\gamma \gamma'}} & **[r] \cH^{\rm spin}_{\gamma}  \ar[d]^{\,^*p^{\rm spin}_{\gamma \gamma'}}
  \\
  \cH_{\gamma'}    \ar[r]^{\cT_{\gamma'}}   & **[r] \cH^{\rm spin}_{\gamma'}
}
\ee
Thus, the equivalence relation $\sim$ on the spinor side\footnote{Here we use the same symbol for equivalence classes at the spinor side as for those on the group side. Essentially these are the same.} can simply be \emph{defined} by just setting two functions $\psi^{\rm spin}_{\gamma}$ and $\phi^{\rm spin}_{\gamma'}$ living on two different graphs to be equivalent, $\psi^{\rm spin}_{\gamma} \sim \phi^{\rm spin}_{\gamma'}$,  if their preimages under the unitary maps $\cT_\gamma$ and $\cT_{\gamma'}$ are:
\be
\psi^{\rm spin}_{\gamma}  \sim  \phi^{\rm spin}_{\gamma'} : \Leftrightarrow \cT^{-1}_\gamma \psi^{\rm spin}_{\gamma} \sim \cT^{-1}_{\gamma'} \phi^{\rm spin}_{\gamma'} \; .\nn
\ee
Thus equivalence classes on the spinor side are simply defined as images under the family of unitary maps $\{\cT_\gamma \}_\gamma$ of equivalence classes on the group side:
\be
\left[  \psi^{\rm spin}_\gamma \right]_{\sim} := \{ \phi_{\gamma'}^{\rm spin}  \in \cH^{\rm spin}_{\gamma'}  | \phi^{\rm spin}_{\gamma'} = \cT_{\gamma'} \circ \,^*p_{\gamma \gamma'} \circ \cT^{-1}_\gamma \psi^{\rm spin}_\gamma    \} \; .\nn
\ee
Defining the continuum spinor Hilbert space as
\be \label{continuum_spinor}
\cH^{\rm spin} := \overline{  \cup_\gamma \cH^{\rm spin}_\gamma / \sim } \; ,
\ee
the family of maps $\{  \cT_\gamma \}_\gamma$ defines a unitary map $\overline{\cT}$ between the continuum Hilbert space of loop quantum gravity and the latter as
\be
\overline{\cT}: \cH \rightarrow \cH^{\rm spin}; [\psi_\gamma] \mapsto \overline{\cT}[\psi_\gamma] := [\cT_\gamma \psi_\gamma] .\nn
\ee
This establishes unitary equivalence between the Hilbert space $\cH = L^2(\overline{\cA}, d\mu_{\rm AL})$ and the continuum spinor space (\ref{continuum_spinor}).\\
For all practical purposes one will use some cutoff in the degrees of freedom and therefore it is mostly enough to restrict oneself to a single graph or a small family thereof. However, from a conceptual point of view it is important to note that the equivalence between the group-based framework and the spinorial framework is not restricted to a fixed graph and/or polyhedral decomposition. The infinite number of degrees of freedom, captured by loop quantum gravity, is also present in the spinor Hilbert space $\cH^{\rm spin}$.\\
However, in order to better understand the equivalence classes on the spinor side one might want to analyze the isometric embeddings $^*p^{\rm spin}_{\gamma \gamma'}$ more intrinsically. In loop quantum gravity these are strongly related to the choice of group variables on the graph and therefore to group multiplication and inversion. As we discussed in section \ref{convolution_sec} the group multiplication property does not hold anymore for the spinor states due to the restriction to the holomorphic part of the group variable. Also, it is an interesting question whether the continuum spinor space $\cH_e^{\rm spin}$ can be written as the space of square integrable functions over some manifold, as is the case for the LQG Hilbert space $\cH$. There, the main reason why this turns out to be true is the fact that the isometric embeddings $\,^*p_{\gamma' \gamma}: \cH_{\gamma'} \rightarrow \cH_\gamma$ can be interpreted as the push-forward of projective maps $p_{\gamma' \gamma}: \cA_\gamma \rightarrow \cA_{\gamma'}$ on the underlying manifolds due to the identification $\cH_\gamma = L^2(\cA_\gamma, d\mu_\gamma)$. Unfortunately this seems not to be case on the spinor side due to the modified multiplication law (\ref{modified_group_law}). These modified multiplication laws will be a good starting point to understand the equivalence classes defined through spinorial variables. We leave this issue open for future research\footnote{As of now the continuum spinor Hilbert space $\cH^{\rm spin}$ is defined via an inductive limit of graph-based Hilbert spaces $\cH_\gamma^{\rm spin}$. Each of these Hilbert spaces describes the quantum geometry of a certain polyhedral decomposition of the 3-manifold. Elements of the continuum spinor Hilbert space are therefore equivalence-classes of such quantum geometries based on infinitely many \emph{different} polyhedral decompositions and thus have a precise interpretation. This is analoguous to the equivalence classes of spin network functions in standard loop quantum gravity as elements of $\cH$. Whether the space $\cH^{\rm spin}$ has an interpretation as the space of square-integrable functions over some functional space describing 3-geometries, analoguous to $\cH = L^2(\bar{\cA}, d\mu)$ in standard loop quantum gravity, is an open question.}.

\subsection{Taking the $\SU(2)$-invariance into account} \label{su2-invariance}
So far we have applied the unitary map $\cT_\gamma$ to \emph{gauge-variant} spin network states, i.e. before taking the $\SU(2)$-invariance at the vertices into account. As the gauge-invariant spin network functions are a true subset of all spin network functions we can apply our map as well to the latter. From the point of view of loop quantum gravity one would simply apply the unitary map $\cT_\gamma$ onto functions constructed out of intertwiners living at the vertices, as these form a basis of the gauge-invariant Hilbert space. However, explicitly working with intertwiners and writing them as suitable combinations of gauge-variant spin network functions is a rather time-consuming and cumbersome task as one has to dive into $\SU(2)$-recoupling theory (which is well understood in principle but rather nasty for explicit computations). Fortunately, using the spinor framework there is a second route to follow which is much less troublesome: considering the spinors as basic variables one can \emph{first} construct $\SU(2)$-invariant functions of spinors and \emph{thereafter} glue these together in a suitable fashion to fulfill the $\U(1)$-invariance on each edge. By doing so one arrives at $\SU(2)$-invariant functions that lie in the image of $\cT_\gamma$ and thus are unitarily related to $\SU(2)$-invariant spin network states. However, in the spinor basis they are much simpler than in the group case as we will see in the following.\\
Implementing $\SU(2)$-invariance before $\U(1)$-invariance has the advantage that this can be done without much effort: For each vertex $v_i$ with valence $N_i$ separately the $F$-variables (see section \ref{F-variables}) parameterize the holomorphic part of the classical $\SU(2)$-invariant phase space associated to that vertex. Therefore polynomials in $F$ are elements of the Bargmann-space $\cF_{2}^{\otimes N_i}$ living at $v_i$ and completely characterize the intertwiner-space of that vertex. The remaining task is then simply to glue these $F$-polynomials of different vertices together in a way that respects the $\U(1)$-invariance at the edges. This leads to certain conditions on the exponents in the polynomials and the result can easily be seen to lie in the image of $\cT^\gamma$, i.e. these polynomials are unitarily equivalent to $\SU(2)$-invariant spin network functions. However, they turn out to be much simpler and therefore open a new door to perform calculations on the gauge invariant level of loop quantum gravity.\\
We will first illustrate the idea on two simple examples, namely the one-loop graph $\gamma^{\rm 1-l}$ (see figure \ref{oneloopgraph}) and the two-vertex graph $\gamma^{2-v}$ (see figure \ref{2vertex}) and thereafter explain how gauge-invariant states can be constructed for an arbitrary graph.

\subsubsection*{One-loop case}
\begin{figure}
\begin{center}
	\includegraphics[scale=.8]{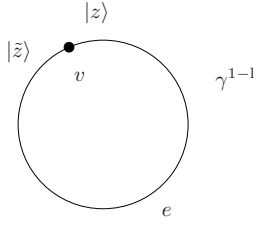}
	\caption{ \label{oneloopgraph} The one-loop graph. Gauge invariant spin network functions on this graph can be mapped unitarily onto polynomials in $F$.}
	\end{center}
\end{figure}

Consider the most elementary graph possible, namely a single vertex connected to itself through one single edge, $\gamma^{\rm 1-l}$ (see figure \ref{oneloopgraph}).
 This graph carries two spinors $\ket{z}, \ket{\tz}$. Using the unitary map $\cT$ on that graph it is clear that a basis of the gauge-variant (edge-)Hilbert space $\cH_e^{\rm spin}$ on $\gamma^{\rm 1-l}$ is given by the spinor coherent states $\cP^j_{\omega \tilde{\omega}}(z,\tz)$ or alternatively in the magnetic number basis by $\cP^j_{mn}(z,\tz)$. What happens if we demand $\SU(2)$-invariance at the vertex? \\
The $\SU(2)$-invariant, holomorphic part of the classical phase space associated to that graph is one-dimensional, the only non-vanishing variable is $F= \braaketrt{z}{\tz}$.
$\SU(2)$-invariant functions on the one loop graph that are holomorphic in both $\ket{z}$ and $\ket{\tz}$ can therefore be written as
\be
f_J(z,\tz) :=  \frac{F^{2J}}{(2J)! \sqrt{d_J}} \, .\nn
\ee
with $J \in \N/2$ and $d_J = 2J+1$.
They are orthonormal with respect to the inner product on $\cF_2 \otimes \cF_2$,
\be
\int d\mu(z) d\mu(\tz) \overline{f_J(z,\tz)}f_K(z,\tz) & = & \frac{\delta_{JK}}{d_J} \, .\nn
\ee
Furthermore, it is easy to see that these functions are already $\U(1)$-invariant and can therefore be written as unitary transforms of spin network functions:
\be
f_J(z,\tz) & = &  \frac{1}{\sqrt{d_J}}\sum\limits_{m=-J}^{+J}\frac{(-z^1)^{J+m}(\tz^0)^{J+m}(z^0)^{J-m}(\tz^1)^{J-m}}{(J+m)!(J-m)!} \nn \\
& = & \frac{1}{\sqrt{d_J}}\sum\limits_{m=-J}^{+J} (-1)^{J+m} \cP^{J}_{-mm}(z, \tz) \nn \\
& = & \cT \left[\frac{1}{\sqrt{d_J}}\sum\limits_{m=-J}^{+J} (-1)^{J+m}  D^{J}_{-m,m}(g)\right](z, \tz)\nn
\ee
To conclude, in the case of the one-loop graph the elements of the gauge- ($\SU(2)$- and $\U(1)$-) invariant Hilbert space are simply given by polynomials in the holomorphic $F$-variables.

\subsubsection*{2-vertex graph}

\begin{figure}
\begin{center}
	\includegraphics[scale=.8]{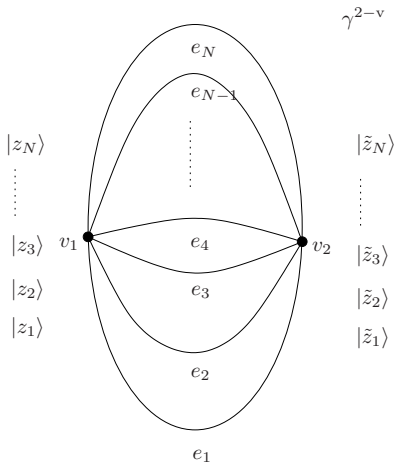}
	\caption{ \label{2vertex} The two vertex graph. Gauge invariant spin network functions on this graph can be mapped unitarily onto traces of polybinomials such as $\Tr(F \tilde{F}), \Tr(F \tilde{F} F \tilde{F}), \dots$}
	\end{center}
\end{figure}

Let us have a look at a slightly more complicated example: the two-vertex graph $\gamma^{\rm 2-v}$ (see figure \ref{2vertex}) which was analyzed in detail in \cite{etera_spanish_un1, etera_spanish_un2}. This graph consists  of two vertices, connected with N edges. The classical phase space associated to that graph consist of $2N$ spinors $\ket{z_a}$ and $\ket{\tz_a}, \; a= 1, \dots, N$ living at vertex $v_1$ and $v_2$ respectively. Gauge-variant spin network states are unitarily related through $\cT$ to elements of the spinor Hilbert space $\cH_{\gamma^{\rm 2-v}}^{\rm spin}$ of the form
\be
\cP^{j_1}_{m_1 n_1}(z_1, \tz_1) \cP^{j_2}_{m_2 n_2}(z_2, \tz_2)\dots \cP^{j_N}_{m_N n_N}(z_N, \tz_N) \, .\nn
\ee
Imposing $\SU(2)$-gauge invariance at both vertices on these states one clearly ends up with the same recoupling theory as when working with standard spin network functions (which is well understood in principle but not very accessible for actual computations). Using the spinor states we can follow the second route described above: \emph{first} impose $\SU(2)$-invariance on the vertices, \emph{thereafter} take care of the $\U(1)$-invariance at the edges. This way one also arrives at states that are unitarily related to gauge-invariant spin network states.\\
The $\SU(2)$-invariant variables of the holomorphic part of the classical phase space are given by
\be
F_{ij} := \braaketrt{z_i}{z_j}, \qquad \tilde{F}_{ij} := \braaketrt{\tz_i}{\tz_j} \, , i,j = 1, \dots, N \; , \nn
\ee
in total $N^2-N$ independent (complex) degrees of freedom (as $F$ and $\tilde{F}$ are both antisymmetric). A general holomorphic, square-integrable function of these variables in the Hilbert space $[\cF_2]^{\otimes2N}$ is given by
\be \label{2v_ansatz}
\prod\limits_{i,j} \left( F_{ij} \right)^{2J_{ij}} \prod\limits_{k,l} \left( \tilde{F}_{kl} \right)^{2\tilde{J}_{kl}} \, ,
\ee
where $J_{ij}, \tilde{J}_{kl} \in \N/2$ and the products run over all values $i,j,k,l = 1, \dots, N$ with $i<j$ and $k<l$. However, these functions are in general not invariant under the $\U(1)$-transformations on the edges as these map $(\ket{z_i}, \ket{\tz_i}) \mapsto (e^{i\phi_i}\ket{z_i}, e^{-i\phi_i}\ket{\tz_i})$. Imposing invariance under $\U(1)^N$ means restricting (\ref{2v_ansatz}) to functions with $J_{ij} = \tilde{J}_{ij}$, i.e. those which contain the same powers of $\ket{z_i}$ and $\ket{\tz_i}$. This can be obtained by simply taking the trace over polybinomials of $F\tilde{F}$, such as
\be
\Tr(F\tilde{F}) = \sum\limits_{i,j = 1}^N F_{ij} \tilde{F}_{ji} \, \nn
\ee
as well as higher orders $\Tr(F\tilde{F}F\tilde{F}), \Tr(F\tilde{F}F\tilde{F}F\tilde{F})$, \dots. These are $(\SU(2))^2$- and $(\U(1))^N$-invariant holomorphic and square-integrable functions of $2N$ spinors and thus elements of\\ $\cH^{\rm spin}_{\gamma^{\rm 2-v}}/(\SU(2))^2$. Their explicit expressions in terms of $\cP^j_{mn}(z_i, \tz_i)$, and thus their relation to standard spin network functions, can easily be computed by rearranging the powers of $\ket{z_i}$ and $\ket{\tz_i}$ appearing in the traces, for example
\be \label{trace_2vertex}
\Tr(F\tilde{F}) & = & \sum\limits_{i,j=1}^N \braaketrt{z_i}{z_j}  \braaketrt{\tz_j}{\tz_i} \\
& = &  \sum\limits_{i,j=1}^N [-z_i^1 z_j^0 + z_i^0 z_j^1][-\tz_j^1 \tz_i^0 + \tz_j^0 \tz_i^1] \nn \\
& = &  \sum\limits_{i,j=1}^N \cP^{\frac{1}{2}}_{-\frac{1}{2} \frac{1}{2}}(z_i,\tz_i)\cP^{\frac{1}{2}}_{\frac{1}{2} -\frac{1}{2}}(z_j,\tz_j) + \cP^{\frac{1}{2}}_{\frac{1}{2} -\frac{1}{2}}(z_i,\tz_i)\cP^{\frac{1}{2}}_{-\frac{1}{2} \frac{1}{2}}(z_j,\tz_j) \nn \\
& & \qquad - \cP^{\frac{1}{2}}_{-\frac{1}{2} -\frac{1}{2}}(z_i,\tz_i)\cP^{\frac{1}{2}}_{\frac{1}{2} \frac{1}{2}}(z_j,\tz_j) - \cP^{\frac{1}{2}}_{\frac{1}{2} \frac{1}{2}}(z_i,\tz_i)\cP^{\frac{1}{2}}_{-\frac{1}{2} -\frac{1}{2}}(z_j,\tz_j) \nn \\
& = & \cT^\gamma \left[  \sum\limits_{i,j=1}^N D^{\frac{1}{2}}_{-\frac{1}{2}, \frac{1}{2}}(g_i)D^{\frac{1}{2}}_{\frac{1}{2}, -\frac{1}{2}}(g_j) + D^{\frac{1}{2}}_{\frac{1}{2}, -\frac{1}{2}}(g_i)D^{\frac{1}{2}}_{-\frac{1}{2}, \frac{1}{2}}(g_j) \right. \nn \\
& & \left. \qquad \qquad - D^{\frac{1}{2}}_{-\frac{1}{2}, -\frac{1}{2}}(g_i)D^{\frac{1}{2}}_{\frac{1}{2}, \frac{1}{2}}(g_j) - D^{\frac{1}{2}}_{\frac{1}{2}, \frac{1}{2}}(g_i)D^{\frac{1}{2}}_{-\frac{1}{2}, -\frac{1}{2}}(g_j) \right](z_1, \dots, z_N, \tz_1, \dots \tz_N) \, . \nn
\ee
Of course, the same result could have been obtained by directly applying the unitary map $\cT_\gamma$ to the gauge invariant spin network functions in the loop quantum gravity Hilbert space $\cH_{\gamma^{\rm 2-v}}$. However, considering the spinors as fundamental variables simplifies the situation drastically: no complicated recoupling theory has to be performed anymore to extract the gauge invariant elements of the Hilbert space and the states can be written as simple polynomials in the $F_{ij}$ and $\tilde{F}_{ij}$. Furthermore, the reformulation of loop quantum gravity on the 2-vertex graph in terms of spinors resembles closely the structure of a matrix model, a route which was investigated already in the articles cited above.

\subsubsection*{Arbitrary graph}

\begin{figure}
\begin{center}
	\includegraphics[scale=.8]{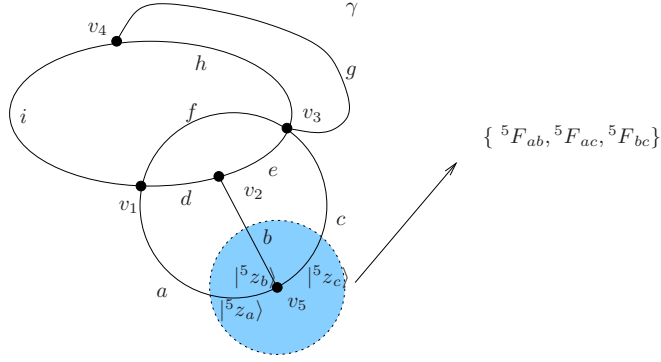}
	\caption{\label{graph} A graph with five vertices ($v_1, v_2, v_3, v_4, v_5$) and nine edges ($a,b,c,d,e,f,g,h,i$). Let $N_i$ be the valence of vertex $v_i$. Then at each vertex the $\SU(2)$-invariant, holomorphic part of the classical phase space is $\frac{N_i^2 - N_i}{2}$-dimensional. For example, at vertex $v_5$ (the highlighted region of the above graph) this space consists of ${^5F}_{ab} = \braaketrt{{^5z}_a}{{^5z}_b}, {^5F}_{ac} = \braaketrt{{^5z}_a}{{^5z}_c}, {^5F}_{bc} = \braaketrt{{^5z}_b}{{^5z}_c}$, where $\ket{{^5z}_a}, \ket{{^5z}_b}, \ket{{^5z}_c}$ are the spinors at vertex $v_5$ into direction of the vertices $a,b,c$ respectively. }
	\end{center}
\end{figure}

\begin{figure}
\begin{center}
	\includegraphics[scale=.8]{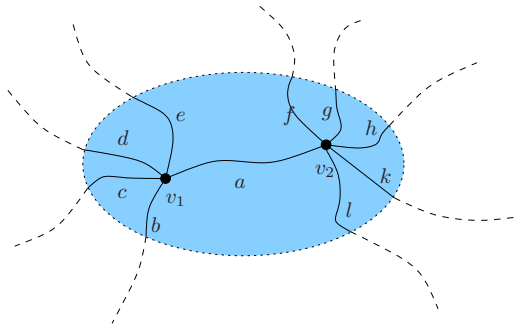}
	\caption{ \label{condition} Taking care of $\U(1)$-invariance on each edge: the highlighted region depicts a part of the total graph $\gamma$ consisting of two vertices $v_1$ and $v_2$ connected by an edge $a$. In order to ensure that (\ref{generic_function}) for this graph is $\U(1)$-invariant one must demand that ${^1J}_{ab} + {^1J}_{ac} + {^1J}_{ad} + {^1J}_{ae} = {^2J}_{af} + {^2J}_{ag} + {^2J}_{ah} + {^2J}_{ak} + {^2J}_{al}$. }
	\end{center}
\end{figure}

\begin{figure}
\begin{center}
	\includegraphics[scale=.8]{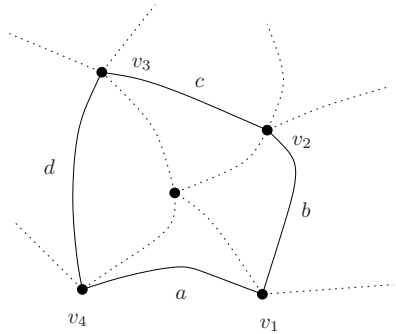}
	\caption{\label{loop_variables} A simple gauge invariant function is given by taking products of $F$-variables along any closed loop within the graph, for instance ${^1F}_{ab} \, {^2F}_{bc}\, {^3F}_{cd}\, {^4F}_{da}$ in this example. These can be shown to be unitarily equivalent to certain combinations of Wigner matrices in the $\frac{1}{2}$-representation of the group on each of the edges involved.  }
	\end{center}
\end{figure}

Now we turn to the situation of an arbitrary, but fixed, graph $\gamma$ with $V$ vertices and $E$ edges and denote by $N_i$ the valence of vertex $v_i$, for an example see figure \ref{graph}. Before imposing either $(\SU(2))^V$- or $(\U(1))^E$-invariance each vertex $v_i$ carries $N_i$ spinors $\{ \ket{{^iz}_1}, \ket{{^iz}_2}, \dots, \ket{{^iz}_{N_i}}   \}$. After demanding $\SU(2)$-invariance (and only allowing holomorphic functions) a minimal set of variables is given by the ${^iF}_{ab} := \braaketrt{{^iz}_a}{{^iz}_b}$ where $a,b$ label two mutually distinct edges connected to the vertex $v_i$. For each vertex the ${^iF}_{ab}$ form a basis of the intertwiner space.\\
However, in order to make contact with the gauge invariant loop quantum gravity Hilbert space, we also have to impose invariance under $(\U(1))^E$ by appropriately glueing together these intertwiner spaces:\\
A general holomorphic function of the $N_\gamma := 2 E = \sum\limits_{i=1}^V N_i$ spinors can be written as
\be \label{generic_function}
\prod\limits_{i = 1}^{V} \prod\limits_{a_i, b_i} \left( {^iF}_{a_i b_i}  \right)^{2 \; {^iJ}_{a_i b_i}} \, ,\nn
\ee
where $a_i, b_i$ label the (ordered) edges connected to vertex $v_i$ and the product runs over all mutually distinct pairs of edges with $a_i < b_i$. ${^iJ}_{a_i b_i} \in \N/2$ are simply the powers with which each $F$ contributes to the total function. These functions\footnote{Note that using the creation operators $^i\hat{F}_{a_i b_i}$ these can simply be written as $\prod\limits_{i = 1}^{V} \prod\limits_{a_i, b_i} \left( {^i\hat{F}}_{a_i b_i}  \right)^{2 \; {^iJ}_{a_i b_i}}\ket{0}$ where the vacuum state is just the unity function. These in turn have been shown to be related to the coherent intertwiners of \cite{eterasimone} in \cite{laurent_etera_un2}.} are holomorphic, square-integrable with respect to the product measure over all spinors $\prod\limits_{i,a} d\mu({^iz_a})$ and invariant under $(\SU(2))^V$. In order to take $\U(1)$-invariance for the edges into account, certain conditions on the ${^iJ}_{a_i b_i}$ have to be fulfilled: (\ref{generic_function}) can only be invariant under $\U(1)$-transformations on the edges if separately for each edge the corresponding spinors on both sides of the edge occur with the same power. This is the case only if (see figure \ref{condition})
\be
\sum\limits_{b} {^iJ}_{ab} = \sum\limits_{c} {^jJ}_{ac} \, ,\nn
\ee
where vertex $v_i$ and $v_j$ are connected by an edge $a$, ${^iJ}_{ab}, {^jJ}_{ac}$ are the corresponding powers in (\ref{generic_function}) and the sums over $b,c$ run over all edges adjacent at vertex $v_i$ or $v_j$ respectively different from $a$.
This leads to $E$ simple conditions on the exponents in (\ref{generic_function}) which -- when solved -- fully characterize the gauge ($\SU(2)$- and $\U(1)$-) invariant kinematical Hilbert space of loop quantum gravity; each function of the prescribed type will be unitarily related to an $\SU(2)$-invariant spin network function through our unitary map $\cT_\gamma$.\\
One particularly easy example of such functions are a certain kind of ``loop variables'': Choose any closed loop within the graph $\gamma$, and simply multiply the $F$-variables along this loop (see figure \ref{loop_variables}). These correspond to certain (sums of) combinations of Wigner-matrices in the $\frac{1}{2}$-representation on each of the edges involved (similar to those in the $2$-vertex example of (\ref{trace_2vertex}) ).

\section{Conclusion}
In this article we have rigorously expanded the recently developed spinorial formulation of loop gravity to the quantum theory. Taking seriously the idea that the fundamental variables are a pair of $\C^2$-spinors per edge and that group- and Lie-algebra-variables are seen as composite variables we are led to a quantization that is different from $\cH_e$, the space of square integrable functions over $\SU(2)$ typically used in loop gravity. Using the spinorial variables, the natural Hilbert space $\cH_e^{\rm spin}$ turned out to be (an appropriate gauge reduction of) the Hilbert space of holomorphic square-integrable functions over two spinors. However, we constructed a unitary map $\cT: \cH_e \rightarrow \cH^{\rm spin}_e$, which from an abstract point of view can be understood as the restriction to the holomorphic part of the group element when the latter is written in terms of spinors. Further, we showed that this map generalizes to an arbitrary graph and is compatible with the inductive limit construction that defines the Hilbert space of the continuum theory. This map shows that the reformulation of loop quantum gravity in terms of spinors captures exactly the same physics.

However, writing loop quantum gravity in holomorphic variables changes the focus and clearly has some advantages compared to the standard framework: at the classical level this reformulation allows for a consistent interpretation in terms of discrete, piecewise flat geometries. Oriented areas of elementary polyhedra are encoded in the $\R^3$-vectors defined through the spinors and the classical phase space can be seen as a phase space of polyhedra glued together in a way such that curvature is generated. This gives the Hilbert space of spin networks, as a quantization of that phase space, an intuitive interpretation as a state space of discrete, piecewise flat geometries.

Besides that we expect that our map will be a very useful computational tool: to date computing physical quantities such as correlation functions in loop quantum gravity is a very difficult task, one of the main reasons (besides conceptual difficulties) being that the calculations involve complicated integrals over $\SU(2)$ which in many cases cannot be solved exactly or even organized in an efficient way for numerical studies. Using the spinor representation for loop quantum gravity we expect to be in a better situation as integrals over $\SU(2)$ are mapped onto straightforward integrals over the complex plane which can be treated using standard tool such as Wick's theorem. We illustrated this by computing the orthonormality of group representation matrices using the fact that the Haar measure on $\SU(2)$ can be written as a product of two Gaussian measures on $\C^2$.

We also expect the results of this paper to be useful in the context of group field theory on $\SU(2)$.
Indeed, we could map the field over the group manifold $\SU(2)$ (or more exactly several copies of $\SU(2)$) to a field over (the corresponding number of copies of) $\C^4$. We expect that this spinor reformulation of group field theory would lead directly to Feynman amplitudes expressed as integrals over spinor variables, which should match the spinfoam amplitudes recently defined in \cite{dupuis_livine_intertwiner2} in term of coherent intertwiners and spinors. As a side-product, this would also provide the group field theory formulation of the new spinfoam model defined in \cite{dupuis_livine_intertwiner2} using the holomorphic simplicity constraints.
Furthermore, we expect the results presented here to be useful in the study of the renormalization properties of group field theories. Indeed, mapping the group field theory action on $\SU(2)^N$ onto a standard field theory action on $\C^{4N}$ could open a new route towards the analysis of divergences in the perturbative expansion of its n-point functions.

Besides that, the reformulation of loop quantum gravity in terms of spinorial variables  bears some similarity to to the recently proposed non-commutative flux representation \cite{flux_representation}. It will be interesting to understand this link better. Finally, it would be enlightening to understand the physical relevance of the spinor variables to investigate the possibility whether this spinorial formulation can be understood as a suitable discretization and quantization of a classical reformulation of gravity in terms of spinor-fields at the continuum level.

\begin{acknowledgments}
This work was partially supported by the ANR ``Programme Blanc'' grant LQG-09. JT was partially supported by the ESF ``Quantum Geometry and Quantum Gravity'' short visit grant 3992.
\end{acknowledgments}

\appendix

\section{Segal-Bargmann transform} \label{segal_bargmann_appendix}
As our construction resembles in some points the Segal-Bargmann transform, we briefly recall the latter in this appendix. The Segal-Bargmann transform \cite{bargmann1} is an invertible, unitary transformation $\cB: \cH_x:= L^2(\R, dx) \rightarrow \cF$ from the space of square-integrable functions over the real line onto the Bargmann space.\\
It can be written as an integral transform with the following integral kernel:
\be
\cK^{\rm SB}(z,x) = \pi^{-1/4} e^{-\frac{1}{2}\bar{z}^2 + \sqrt{2}\bar{z}x - \frac{1}{2}x^2 } \; .\nn
\ee
The Segal-Bargmann transform is then given by:
\be
\cB & : & \cH_x \rightarrow \cF; f \mapsto \cB f \nn \\
& & (\cB f)(z) := \int\limits_\R dx \overline{\cK^{\rm SB}(z,x)} f(x) \nn
\ee
or
\be
(\cB f)(z)= \braaket{\cK^{\rm SB}_z}{f}_{\cH_x}\nn
\ee
when interpreting the kernel $\K^{\rm SB}_z(x) := \cK^{\rm SB}(z,x)$ as simply a function of $x$ and denoting by $\braaket{\cdot}{\cdot}_{\cH_x}$ the inner product on $\cH_x$.\\
By also defining $\cK^{\rm SB}_x(z) := \cK^{\rm SB}(z,x)$ interpreted as a function of $z$ one can easily check that the transform is in fact unitary by showing that
\be
\braaket{\cK^{\rm SB}_z}{\cK^{\rm SB}_{z'}}_{\cH_x} = \delta_{z'}(z) \; , \nn \\
\braaket{\cK^{\rm SB}_x}{\cK^{\rm SB}_{x'}}_{\cF} = \delta_{x'}(x) \; ,\nn
\ee
where the second one is just the standard delta-distribution in $L^2(\R, dx)$ and the first one is given by $e^{z'\bar{z}}$.
The inverse Segal-Bargmann transform is then given as a map
\be
\cB^{-1} & : & \cF \rightarrow \cH_x \; ; \nn \\
& & \tilde{f} \mapsto (\cB^{-1}\tilde{f})(x) := \braaket{\overline{\cK^{\rm SB}_x}}{\tilde{f}}_{\cF} \; .\nn
\ee
This isometric isomorphisms can be made more precise by analyzing how the elements of a given basis are transformed. The inverse Segal-Bargmann transform maps the holomorphic polynomials $e_n(z) := \frac{1}{\sqrt{n!}}z^n$ onto the Hermite functions of degree $n$, i.e. solutions to the differential equation\footnote{The Hermite functions are the ``weighted'' and normalized versions of the Hermite polynomials. They are true elements of $L^2(\R, dx)$ with just the Lebesgue measure $dx$. They differ from the commonly used Hermite polynomials by a factor of $e^{-\frac{1}{2}x^2}$ and by some normalization constant. They form a complete orthonormal basis of $L^2(\R, dx)$ with completeness relation $\sum\limits_{n \in \N} H_n(x) H_n(x') = \delta(x,x')$.  }.
\be
(-\partial_x^2 + x^2 - 2n -1)H_n(x) = 0 \; ,\nn
\ee
thus,
\be
(\cB H_n)(z) = e_n(z) \; .\nn
\ee
It is easy to show, that -- as expected -- one gets the following correspondence between elementary operators on both spaces
\be
z & \leftrightarrow & \frac{1}{\sqrt{2}}(x - \partial_x) \nn \; , \\
\partial_z &\leftrightarrow&  \frac{1}{\sqrt{2}}(x + \partial_x) \nn \; .
\ee

\section{Coherent state basis for $\SU(2)$} \label{coherent_state_appendix}
In this appendix we review the coherent state basis for $\SU(2)$ and collect some useful definitions and formulae used in the main text.\\
Consider the $d_j := 2j+1$ dimensional vector space $\cV^j$ with basis $\ket{j,m}$ where $j \in \N/2$ and $m= -j, \dots, +j$. The basis is orthonormal and complete
\be
\braaket{j,m}{j', m'} = \delta^{jj'}\delta_{mm'}, \qquad \one_j = \sum\limits_m \ket{j,m}\bra{j,m} \, ,\nn
\ee
and is commonly referred to as the \emph{magnetic number basis}.
Each $\V^j$ carries an irreducible unitary representation of $\SU(2)$ and the matrix elements of that representation in the magnetic number basis are denoted by $D^j_{mn}(g) := \braaaket{j,m}{D(g)}{j,n}$ and, being representation matrices, fulfill the ``group law''
\be \label{group_multiplication}
\sum\limits_k D^j_{mk}(g_1) D^j_{kn}(g_2) = D^j_{mn}(g_1 g_2) \, .
\ee
Choosing a coordinate system on $\SU(2)$, for example
\be
g = \begin{pmatrix} \alpha & \beta \\ -\bar{\beta} & \bar{\alpha} \end{pmatrix}, \qquad |\alpha|^2 + |\beta|^2 = 1\nn
\ee
one can give an explicit formula for the $D^j_{mn}(g)$:
\be \label{rep_mat}
D^j_{mn}(g) & = & \sum\limits_{k=0}^{j+n}\sum\limits_{l=0}^{j-n}\frac{ \sqrt{ (j+n)!(j-n)!(2j-k-l)!(k+l)!}  }{  k!l!(j+n-k)!(j-n-l)!  } \delta_{m, j - (k+l)} \alpha^{j+n-k} \bar{\alpha}^l \beta^{j-n-l}(-\bar{\beta})^k \, .
\ee
These representation matrices are of particular importance because, according to the Peter-Weyl theorem, they form a complete, orthogonal basis of $L^2(\SU(2), dg)$, the space of square-integrable functions over $\SU(2)$ with respect to the Haar measure $dg$:
\be
\int dg \overline{D^j_{mn}(g)} D^{j'}_{m'n'}(g) & = & \frac{\delta^{jj'}}{d_j}\delta_{mm'}\delta_{nn'} \nn \\
\sum\limits_{j \in \N/2} \sum\limits_{m,n = -j}^{+j} d_j \overline{D^j_{mn}(g')}D^j_{mn}(g) & = & \delta_{g'}(g) \, .\nn
\ee
Thus, every function $f \in L^2(\SU(2))$ can be decomposed into representation matrices as
\be \label{PW}
f(g) & = & \sum\limits_{j,m,n} \sqrt{d_j}\hat{f}^j_{mn}D^j_{mn}(g) \nn \\
\hat{f}^j_{mn} & := & \int dg \sqrt{d_j} \overline{D^j_{mn}(g)} f(g) \, .
\ee
As the explicit expressions  of $D^j_{mn}(g)$ in the magnetic number basis are rather ugly it is useful to introduce a second basis of $\cV^j$, the so called coherent state basis. Choose a spinor $\ket{\omega} := \vectwo{\omega^0}{\omega^1} \in \C^2$ and define
\be
\ket{j,\omega} := \sum\limits_{m=-j}^{+j} \frac{\sqrt{(2j)!}(\omega^0)^{j+m}(\omega^1)^{j-m}}{\sqrt{(j+m)!(j-m)!}}\ket{j,m} \, .\nn
\ee
These states are normalized as $\braaket{j,\omega}{j',\tilde{\omega}} = \delta^{ij} \braaket{\omega}{\tilde{\omega}}^{2j}$ and the identity on $\cV^j$ in terms of coherent states reads
\be
\one_j = \frac{1}{(2j)!}\int d\mu(\omega) \ket{j,\omega}\bra{j,\omega} \, ,\nn
\ee
where $d\mu(\omega) = \frac{1}{\pi^2}e^{-\braaket{\omega}{\omega}}d\omega^0 d\omega^1$ is the normalized Gaussian measure on $\C^2$. They fulfill the nice property that
\be
D^j(g)\ket{j, \omega} = \ket{j,g\omega}\nn
\ee
and factorize into
\be \label{factorization_property}
\ket{j,\omega} = \ket{\frac{1}{2}, \omega}^{\otimes 2j} = \ket{z}^{\otimes 2j} \, .
\ee
The representation matrices of $\SU(2)$ in the coherent state basis can therefore be written as
\be
D^j_{\omega \tilde{\omega}}(g) := \braaaket{j,\omega}{D(g)}{j, \tilde{\omega}} = \sum\limits_{m,n= -j}^{+j} \frac{(2j)!(\bar{\omega}^0)^{j+m}(\bar{\omega}^1)^{j-m}(\tilde{\omega}^0)^{j+n}(\tilde{\omega}^1)^{j-n}}{\sqrt{(j+m)!(j-m)!(j+n)!(j-n)!}} D^j_{mn}(g) \, ,\nn
\ee
or alternatively, using the factorization-property (\ref{factorization_property}) as
\be \label{braket_states}
D^j_{\omega \tilde{\omega}}(g) = \braaaket{\omega}{g}{\tilde{\omega}}^{2j} \, ,
\ee
which is a very useful expression for many computations.\\
The completeness relations in terms of the coherent state basis read as
\be \label{completeness_coherent_group}
\int \overline{D^j_{\omega \tilde{\omega}}(g)} D^{j'}_{\alpha \tilde{\alpha}}(g) & = & \frac{\delta^{jj'}}{d_j}\braaket{\alpha}{\omega}^{2j}\braaket{\tilde{\omega}}{\tilde{\alpha}}^{2j} \nn \\
\sum\limits_j \int d\mu(\omega) \int  d\mu(\tilde{\omega}) \frac{d_j}{[(2j)!]^2} \overline{D^j_{\omega \tilde{\omega}}(g')} D^j_{\omega \tilde{\omega}}(g) & = & \delta_{g'}(g) \, ,
\ee
or using (\ref{braket_states}),
\be
\int dg \braaaket{\tilde{\omega}}{g^{-1}}{\omega}^{2j} \braaaket{\alpha}{g}{\tilde{\alpha}}^{2j'} & = & \frac{\delta^{jj'}}{d_j} \braaket{\alpha}{\omega}^{2j} \braaket{\tilde{\omega}}{\tilde{\alpha}}^{2j} \nn \\
\sum\limits_{j} \int d\mu(\omega)  \frac{d_j}{(2j)!} \braaaket{\omega}{{g'}^{-1}g}{\omega}^{2j} & = & \delta_{g'}(g) \, .
\ee
Furthermore, the group law (\ref{group_multiplication}) in terms of coherent states can be written as
\be \label{group_mult_coherent}
\int d\mu(z) \braaaket{\omega}{g_1}{z}^{2j} \braaaket{z}{g_2}{\tilde{\omega}}^{2j} & = & (2j)!\braaaket{\omega}{g_1 g_2}{\tilde{\omega}}^{2j} \, .
\ee
This leads to the following expression for the Peter-Weyl theorem in terms of coherent states:
\be
f(g) & = & \sum\limits_j \int d\mu(\omega)\int d\mu(\tilde{\omega})\frac{\sqrt{d_j}}{(2j)!}  \check{f}^j_{\omega \tilde{\omega}} \braaaket{\omega}{g}{\tilde{\omega}}^{2j} \nn \\
\check{f}^j_{\omega \tilde{\omega}} & := & \int dg \frac{\sqrt{d_j}}{(2j)!} \braaaket{\tilde{\omega}}{g^{-1}}{\omega}^{2j} f(g) \, .\nn
\ee

\section{LQG and projective techniques} \label{lqg}

In this appendix we briefly collect some facts concerning loop quantum gravity as continuum theory as a supplement for section \ref{cyl_cons} where we have shown that the spinor Hilbert space respects the conditions of cylindrical consistency and thus is capable of carrying the infinite number of degrees of freedom of the gravitational field. For a complete account we refer the reader to \cite{thiemannbook}.\\
It is sometimes stated that the (gauge variant) Hilbert space of loop quantum gravity is given by $\cH_\gamma := L^2(\SU(2)^E, dg)$, where $E$ is the number of edges of a prescribed graph $\gamma$ and $dg$ is the Haar measure on $\SU(2)$.  However, one should keep in mind that each space $\cH_\gamma$ carries only a finite number of degrees of freedom, and in general it is difficult to tell which degrees of freedom these are. Taking two Hilbert spaces $\cH_\gamma$ and $\cH_{\gamma'}$ associated to different graphs $\gamma \neq \gamma'$, what is the relation between two given states $\psi_\gamma \in \cH_\gamma$ and $\phi_{\gamma'} \in \cH_{\gamma'}$? Is there any overlap in the physical information captured by these two finite dimensional Hilbert spaces?\\
This issue is accounted for in LQG by carefully analyzing which states $\psi_\gamma$ and $\phi_{\gamma'}$ are to be considered as equivalent, $\psi_\gamma \sim \phi_{\gamma'}$, despite them being defined on different graphs. Thus, elements of the LQG Hilbert space are \emph{not} states living on a single graph, but equivalence classes of states $[\psi_\gamma]_\sim := \{ \phi_{\gamma'} \in \cH_{\gamma'}: \cH_{\gamma} \ni \phi_{\gamma'} \sim \psi_{\gamma}  \}$ and the LQG Hilbert space is a space of such equivalence classes:
\be
\cH := \overline{\cup_\gamma \cH_\gamma / \sim} \; .\nn
\ee
The space $\cH$ describes the infinite number of degrees of freedom of the gravitational field.\\
Let us briefly describe this space and its elements to understand the equivalence relation imposed on states living on different graphs: The configuration variables used in LQG are the holonomies of the Ashtekar connection along semi-analytic, embedded edges $h_e[A]$. These are elements of $\SU(2)$ with the properties
\be
h_{e_1 \circ e_2}[A] & = & h_{e_1}[A] h_{e_2}[A] \; , \nn \\
h_{e^{-1}}[A] & = & \left( h_e[A]  \right)^{-1} \; . \nn
\ee
Thus, the connection $A$ can be interpreted as an element of $\Hom(\cP, \SU(2))$, namely homomorphism from the groupoid of paths $\cP$ into the gauge group $\SU(2)$. The classical configuration space $\cA$ of $\SU(2)$-connections is clearly a subspace of $\Hom(\cP, \SU(2))$. However, the latter is much bigger because it does not respect the differentiability or continuity present in $\cA$. However, in the context of loop quantum gravity it turns out that $\bar{\cA} :=\Hom(\cP, \SU(2))$ plays the role of the ''quantum configuration space'' over which the field theory Hilbert space is constructed as $\cH = L^2(\bar{\cA}, d\mu_{\rm AL})$ with an appropriate measure $\mu_{\rm AL}$. Proving the equality $L^2(\bar{\cA}, d\mu_{\rm AL}) = \overline{\cup_\gamma \cH_\gamma / \sim}$ is slightly involved, for details see the reference cited above.\\
 However, this equality allows to view the Hilbert space $\cH$ from two different points of view: On the one hand as the space of square integrable functions over the space $\bar{\cA}$ of generalized connections, where $\bar{\cA}$ can be shown to arise as the \emph{projective limit} of a \emph{projective family} of graph based configuration spaces $\cA_\gamma$. On the other hand as an \emph{inductive limit} of an \emph{inductive family} of graph based Hilbert spaces $\cH_\gamma := L^2(\cA_\gamma, d\mu_\gamma)$.
 The natural projective maps\footnote{A graph $\gamma$ is called bigger than a graph $\gamma'$, $\gamma \geq \gamma'$ if every edge $e' \in \gamma'$ can be written as a combination of edges (and inverse edges) $e \in \gamma$. $\geq$ is a partial ordering and the set of all graphs is a partially ordered, directed set.} $p_{\gamma' \gamma}: \cA_\gamma \rightarrow \cA_{\gamma'} \quad \forall \gamma \geq \gamma'$ give rise to natural isometric embeddings $\,^*p_{\gamma'\gamma}: \cH_{\gamma'} \rightarrow \cH_\gamma \quad \forall \gamma \geq \gamma'$ into the other direction by push-forward. As these maps are essential to understand the equivalence classes $[\psi_\gamma]_\sim$ in $\cH$, let us explain them a bit more detailed:\\
For each given graph $\gamma$ the configuration space $\cA_\gamma$ is isomorphic to $(\SU(2))^E$ where $E$ is the number of edges of $\gamma$. An element $A_\gamma \in \cA_\gamma$ can therefore always be identified with an $E$-tuple of $\SU(2)$-elements. Thus, for any $\gamma' \leq \gamma$ there exist a natural projection $p_{\gamma' \gamma}: \cA_\gamma \rightarrow \cA_{\gamma'}$, because the algebraic structure in $\cP$ is directly linked with group multiplication (and inversion) of elements in $\SU(2)$. The set $\{ \cA_\gamma, p_{\gamma' \gamma} \; \forall \gamma \geq \gamma' \}$ is a projective family and and it turns out that this structure (together with the compactness of $\SU(2)$) is sufficient to equip the projective limit $\bar{\cA}$ with  topology and  measure, which is needed to construct the Hilbert space $L^2(\bar{\cA}, d\mu_{\rm AL})$.\\
On the other hand, on the level of Hilbert spaces the situation is the following: to each graph $\gamma$ one can assign a Hilbert space $\cH_\gamma := L^2(\cA_\gamma, d\mu_\gamma)$ which by the isomorphism $\cA_\gamma \simeq (\SU(2))^E$ is unitarily equivalent to $\cH_\gamma \simeq L^2((\SU(2))^E, dg)$. The projective maps $p_{\gamma' \gamma}$ can be pushed forward to the level of Hilbert spaces and thus define injections $\,^*p_{\gamma' \gamma}: \cH_{\gamma'} \rightarrow \cH_\gamma \quad \forall \gamma \geq \gamma'$. Because the family of measures $\{ \mu_\gamma \}$ is cylindrically consistent, $\,^*p_{\gamma' \gamma}$ turn out to be \emph{isometric embeddings}, i.e.
\be
\braaket{\,^*p_{\gamma' \gamma} \psi_{\gamma'}}{  \,^*p_{\gamma' \gamma} \phi_{\gamma'}   }_{\cH_\gamma} = \braaket{ \psi_{\gamma'}  }{  \phi_{\gamma'}   }_{\cH_{\gamma'}} \quad \forall \phi_{\gamma'}, \psi_{\gamma'} \in \cH_{\gamma'}\nn
\ee
This gives a natural notion of equivalence between states defined on different graphs,
\be
\psi_\gamma \sim \phi_{\gamma'} \; : \Leftrightarrow \; \exists \,^*p_{\gamma' \gamma} \mbox{ s.t. } \psi_{\gamma} = \,^*p_{\gamma' \gamma} \phi_{\gamma'} \; ,\nn
\ee
and one can define equivalence classes as
\be
[\psi_{\gamma_1}]_\sim := \{ \phi_{\gamma_2} \in \cH_{\gamma_2} | \phi_{\gamma_2} = \,^*p_{\gamma_1 \gamma_2} \psi_{\gamma_1}   \} \; .\nn
\ee
These are the elements of the loop quantum gravity Hilbert space $\cH$. In particular, as can be seen from the last equation, given any function $\psi_{\gamma_1} \in \cH_{\gamma_1}$ on a given graph $\gamma_1$, one directly can identify an equivalent representative on \emph{any bigger graph $\gamma_2 \geq \gamma_1$}. Into the other direction this is not true: Going from bigger to smaller graphs one inevitably looses information, which means that one cannot identify an equivalent representative on a smaller graph for every function. This shows that the inductive limit construction and the associated techniques of cylindrical consistency are well adapted to construct the ''ultraviolet limit'' of the theory, i.e. the regime of finer and finer graphs which corresponds to the Planck-regime of the theory. However, the ''infrared regime'', i.e. the question how to dynamically derive low energy physics by an appropriate averaging over high energy degrees of freedom, cannot be treated with these methods. One will have to understand how to deal with the lost information in going from bigger to smaller graphs.

\section{Alternative measures on $\C^2$ and the factor of $\sqrt{d_j}$} \label{factor_dj_appendix}

\subsection*{Alternative measures}
In section \ref{quantum_spinors} we worked with a Gaussian measure $d\mu(z) d\mu(\tz)$ on $\C^2 \times \C^2$ which results in the completeness relations (\ref{completeness_coherent_spinors}). Polynomials in the spinors are orthogonal with respect to that measure, therefore in order to get the factor of $\frac{1}{d_j}$ which is present in the completeness relations for $\SU(2)$-coherent states (see formula (\ref{completeness_coherent_group})) the basis states on the spinor side must be weighted with a factor of $\frac{1}{\sqrt{d_j}}$ in order to make the map $\cT$ unitary. This factor, introduced by hand in our construction, can be re-absorbed in a change of measure as we will explain in the following:
Consider the Gaussian measure $d\mu(z) = \frac{1}{\pi^2} dz e^{-\braaket{z}{z}}$ on $\C^2$ and an integral similar to the ones that occur in the completeness relations (ignoring the integral over $\ket{\tz}$ for the moment). A factor of $d_j$ can easily be absorbed in the measure:
\be
&& d_j \frac{1}{\pi^2} \int dz e^{-\braaket{z}{z}} \frac{1}{(2j)!} \braaket{\omega}{z}^{2j} \braaket{z}{\tilde{\omega}}^{2j} \nn \\
& = & \frac{1}{\pi^2} \int dz e^{-\braaket{z}{z}} \frac{1}{(2j)!} \left( \ket{z}\cdot\partial_{\ket{z}} + 1\right) \braaket{\omega}{z}^{2j} \braaket{z}{\tilde{\omega}}^{2j} \nn \\
& = & \frac{1}{\pi^2} \int dz \left[  \left(-\ket{z} \cdot \partial_{\ket{z}} - 1 \right) e^{-\braaket{z}{z}}  \right]\frac{1}{(2j)!} \braaket{\omega}{z}^{2j} \braaket{z}{\tilde{\omega}}^{2j} \nn \\
& = & \frac{1}{\pi^2} \int dz (\braaket{z}{z} - 1) e^{-\braaket{z}{z}} \frac{1}{(2j)!} \braaket{\omega}{z}^{2j} \braaket{z}{\tilde{\omega}}^{2j} \nn \\
& =: & \int d\mu^+(z) \frac{1}{(2j)!} \braaket{\omega}{z}^{2j} \braaket{z}{\tilde{\omega}}^{2j}\nn
\ee
Alternatively, one can follow the logic of section \ref{haar_measure} to see that including certain polynomials of $\braaket{z}{z}$ in the integral just amounts to a change by some combinatorial factor. In the case of the polynomial $(\braaket{z}{z}-1)$ this combinatorial factor turns out to be just $d_j$.\\
Following the same logic, one can show that a factor of $\frac{1}{d_j}$ can be absorbed into the integral by changing the measure to $d\mu^-(z) := \frac{1}{\pi^2}dz \frac{1}{\braaket{z}{z}}e^{-\braaket{z}{z}}$.\\
Using this modified measure, the factor of $\frac{1}{\sqrt{d_j}}$ in the basis elements $\cP^j_{\omega \tilde{\omega}}(z, \tz) = \frac{1}{(2j)! \sqrt{d_j}} \braaket{\omega}{z}^{2j} \brar{\tz} \epsilon \ket{\tilde{\omega}}^{2j}$ is not necessary anymore and one obtains directly
\be
\int d\mu^-(z) \int d\mu(\tz) \frac{1}{[(2j)!]^2} \overline{ \braaket{\omega}{z}^{2j} \brar{\tz} \epsilon \ket{\tilde{\omega}}^{2j} } \braaket{\alpha}{z}^{2k} \brar{\tz} \epsilon \ket{\tilde{\alpha}}^{2k} & = & \frac{\delta^{jk}}{d_j} \braaket{\tilde{\omega}}{\tilde{\alpha}}^{2j} \braaket{\alpha}{\omega}^{2j} \, .\nn
\ee
However, the price to pay is that the new measure on $\C^2 \times \C^2$ is not symmetric under an exchange $\ket{z} \leftrightarrow \ket{\tz}$ anymore.\\
Choosing a new measure also changes the kernel $\cK_g(z,\tz)$ that generates the unitary transform $\cT$: Instead of $\cK_g(z, \tz) = \sum\limits_{k=0}^\infty \frac{\sqrt{d_j}}{k!} \brar{\tz}\epsilon g^{-1} \ket{z}^k$ one now obtains $\cK^-_g(z, \tz) := \sum\limits_{k=0}^\infty \frac{d_j}{k!} \brar{\tz}\epsilon g^{-1} \ket{z}^k = (\brar{\tz}\epsilon g^{-1} \ket{z} + 1) e^{\brar{\tz}\epsilon g^{-1} \ket{z}}$.\\
In fact, there is some arbitrariness in the choice of measure, basis states on the spinor space and integration kernel. One just needs to be careful to make a consistent choice. Here we list the three most convenient ones:
\begin{itemize}
\item Standard Gaussian measure $d\mu(z)d\mu(\tilde{z})$ and integration kernel given by\\ $\cK_g(z, \tz) = \sum\limits_{k=0}^\infty \frac{\sqrt{d_j}}{k!} \brar{\tz}\epsilon g^{-1} \ket{z}^k$: This maps the spin network functions onto orthogonal polynomials weighted by a factor of $\frac{1}{\sqrt{d_j}}$. This seems to be the most natural choice in terms of measure. However, the factor $\frac{1}{\sqrt{d_j}}$ seems a bit ad hoc and the integral kernel has no nice closed expression.
\item Measure with a negative weight $d\mu^-(z)d\mu(z) := \frac{1}{\braaket{z}{z}}d\mu(z)d\mu(\tz)$ and integration kernel given by $\cK^-_g(z,\tz) = (\brar{\tz}\epsilon g^{-1} \ket{z} + 1) e^{\brar{\tz}\epsilon g^{-1} \ket{z}} $: This maps the spin network functions onto orthogonal polynomials without any weight factor. The integration kernel is considerably nicer than in the last example but the measure is not symmetric anymore.
\item Measure with positive weight $d\mu^+(z)d\mu(\tz) := (\braaket{z}{z} - 1) d\mu(z) d\mu(\tz)$ and integration kernel $\cK^+_g(z,\tz) := e^{\brar{\tz}\epsilon g^{-1} \ket{z}}$: This maps spin network functions onto orthonormal polynomials weighted by a factor of $\frac{1}{d_j}$. The integration kernel looks most natural in this choice, however the measure is still non-symmetric.
\end{itemize}
In fact, one can find an expression for an arbitrary (positive) power of $d_j$ in the integral by writing
\be
&& d_j^N \frac{1}{\pi^2}\int dz e^{-\braaket{z}{z} } \braaket{\omega}{z}^{2j} \braaket{z}{\tilde{\omega}}^{2j} \nn \\
& = &  \frac{1}{\pi^2}\int dz e^{-\braaket{z}{z} } (\ket{z}\cdot \partial_{\ket{z}} + 1)^N\braaket{\omega}{z}^{2j} \braaket{z}{\tilde{\omega}}^{2j} \nn \\
& = &  \frac{1}{\pi^2}\int dz  \left[ (-\ket{z}\cdot \partial_{\ket{z}} - 1)^Ne^{-\braaket{z}{z} } \right] \braaket{\omega}{z}^{2j} \braaket{z}{\tilde{\omega}}^{2j} \nn \\
& = & \frac{1}{\pi^2}\int dz  P_N(\braaket{z}{z}) e^{-\braaket{z}{z} } \braaket{\omega}{z}^{2j} \braaket{z}{\tilde{\omega}}^{2j} \, ,\nn
\ee
where $P_N(x)$ is a class of polynomials generated by
\be
P_N(x) = e^{x}(-x\partial_x - 1)^N e^{-x} \, .\nn
\ee
They can be shown to fulfill the recursion relations
\be
P_N(x) = (x-1 - x\partial_x)P_{N-1}(x), \quad P_0(x) = 1\, .\nn
\ee
A general class of measures on $\C^2$ can then be defined as
\be
d_N\mu(z) := \frac{1}{\pi^2} dz P_N(\braaket{z}{z}) e^{-\braaket{z}{z}} \nn
\ee
Each measure $d_N\mu(z)$ is constructed such that weighted polynomials $d_j^{-\frac{N}{2}}e^j_m(z) = \frac{(z^0)^{j+m}(z^1)^{j-m}}{\sqrt{d^N_j (j+m)!(j-m)!}}$ are orthogonal:
\be
\int d_N\mu(z) \overline{d_j^{-\frac{N}{2}} e^j_m(z) }d_j^{-\frac{N}{2}} e^k_n(z) = \delta^{jk} \delta_{mn} \nn
\ee

\subsection*{Combinatorial origin of the factor $\sqrt{d_j}$}
In section \ref{sec:unitary_map} we saw that our unitary transform can be understood as reducing the representation matrices of $\SU(2)$ to their holomorphic part as in (\ref{restriction_hol}). It turns out that the factor $\frac{1}{\sqrt{d_j}(2j)!}$ in that map can be understood from exactly this perspective and the change of combinatorics when reducing the full representation matrices $D^j_{\omega \tilde{\omega}}$ to their holomorphic parts.\\
Let us go back to the argument of section \ref{haar_measure} where we showed that the Haar measure on $\SU(2)$ can be written simply as a product of two Gaussian measures on $\C^2$ when considering the spinors $(\ket{z}, \ket{\tz})$ as fundamental variables, and the group element as composite.\\
Rewriting the completeness relation for two $\SU(2)$-representation matrices using the holomorphic-antiholomorphic splitting of the group element we get
\be
&& \int dg \overline{D^j_{\omega \tilde{\omega}}(g)} D^k_{\alpha \tilde{\alpha}}(g) \nn \\
& = & \int d\mu(z) d\mu(\tz) \frac{\left[ - \braaket{\tilde{\omega}}{\tilde{z}}\braaketrt{z}{\omega} + \braakettr{\tilde{\omega}}{\tz} \braaket{z}{\omega}  \right]^{2j}  \left[  \braaket{\alpha}{z}\braaketrt{\tilde{z}}{\tilde{\alpha}}  - \braakettr{\alpha}{z} \braaket{\tz}{\tilde{\alpha}}   \right]^{2k}}{\braaket{z}{z}^{j+k}\braaket{\tz}{\tz}^{j+k}} \, . \nn
\ee
Restricting the group elements to their holomorphic part means throwing away both terms with the minus signs in the above formula, thus we get the following integral:
\be
&& \int dg \overline{D^j_{\omega \tilde{\omega}}(g|_{\rm hol})} D^k_{\alpha \tilde{\alpha}}(g|_{\rm hol}) \nn \\
& = & \left[ \frac{C(0)}{C(j+k)} \right]^2   \int d\mu(z) d\mu(\tz) \left[ \braakettr{\tilde{\omega}}{\tz} \braaket{z}{\omega}  \right]^{2j}  \left[  \braaket{\alpha}{z}\braaketrt{\tilde{z}}{\tilde{\alpha}}  \right]^{2k} \nn \\
& = & \left[ \frac{C(0)}{C(j+k)} \right]^2 \delta^{jk} [(2j)!]^2 \braaket{\alpha}{\omega}^{2j}\braaket{\tilde{\omega}}{\tilde{\alpha}}^{2j} \nn \\
& = & \frac{1}{d^2_j} \delta^{jk} \braaket{\alpha}{\omega}^{2j}\braaket{\tilde{\omega}}{\tilde{\alpha}}^{2j} \nn
\ee
The second to last line can either be computed using the group-multiplication property for $\SU(2)$-coherent states or alternatively, to stay closer to the original derivation of section \ref{haar_measure}, in terms of Wick's theorem: erasing the antiholomorphic part simply changes the number of possible pairings. Thus, if one wants the holomorphic restriction of the representation matrices to have the same orthogonality properties as the full ones (with respect to the same measure), this can only be achieved by ``weighting'' them with a combinatorial factor: the $\frac{1}{(2j)!}$ comes from the fact that we disregarded the norms in the denominator, and the $\frac{1}{\sqrt{dj}}$ from disregarding the antiholomorphic part of the numerator .

\providecommand{\href}[2]{#2}\begingroup\raggedright\endgroup

\end{document}